\documentclass[twocolumn]{aa}

\usepackage{graphicx} 
\usepackage[figuresright]{rotating}
\usepackage{txfonts} 
\begin{document}

   \title{Mg abundances in metal-poor halo stars as a tracer of early Galactic mixing}

   \author{E. Arnone
          \inst{1,2}, 
          S. G. Ryan\inst{1},
          D. Argast\inst{3},
          J. E. Norris\inst{4},
          \and T. C. Beers\inst{5}
          }

   \offprints{S.G. Ryan}

   \institute{Dept. of Physics and Astronomy, The Open University, Milton Keynes, MK7 6AA, U.K.
              \email{s.g.ryan@open.ac.uk}
         \and
             Dept. of Physics and Astronomy, University of Leicester, Leicester, LE1 7RH, U.K.
              \email{e.arnone@ion.le.ac.uk}
         \and
             Dept. of Physics and Astronomy, University of Basel, Basel, Switzerland
              \email{argast@astro.unibas.ch}
         \and
             Research School of Astronomy and Astrophysics, The Australian National University, Weston Creek ACT 2611, Australia
              \email{jen@mso.anu.edu.au}
         \and
             Dept. of Physics and Astronomy and JINA: Joint Institute for Nuclear Astrophysics, Michigan State University, East Lansing, MI 48824
              \email{beers@pa.msu.edu}}

   \date{Received 06/04/04; accepted 27/09/04}

\abstract{ We present results of a detailed chemical analysis performed on
23 main-sequence turnoff stars having $-3.4\leq$[Fe/H]$\leq-2.2$, a sample
selected to be highly homogeneous in T$_{\rm eff}$ and log($g$). We investigate
the efficiency of mixing in the early Galaxy by means of the [Mg/Fe] ratio, and
find that all values lie within a total range of 0.2 dex, with a standard
deviation about the mean of 0.06 dex, consistent with measurement errors. This
implies there is {\it little or no intrinsic scatter} in the early ISM, as
suggested also by the most recent results from high-quality VLT observations.
These results are in contrast with inhomogeneous Galactic chemical evolution
(iGCE) models adopting present supernova (SN) II yields, which predict a peak-to-peak scatter in [Mg/Fe] as high as 1 dex
at very low metallicity, with a corresponding standard deviation of about 0.4
dex. We propose that cooling and mixing timescales should be investigated in
iGCE models to account for the apparent disagreement with present observations. The contrast between the constancy and small dispersion of [Mg/Fe]
reported here and the quite different behaviour of [Ba/Fe] indicates,
according to this interpretation, that Mg and Ba are predominantly
synthesised in different progenitor mass ranges.

   }
   \titlerunning{Mg in metal-poor stars}
   \authorrunning{Arnone {\it et al.} }
   \maketitle
%
\section{Introduction}

The chemical evolution of the Milky Way arises from the
continuous exchange of material between stars and the interstellar medium (ISM). The surfaces of long-lived, low-mass stars retain the composition of the
ISM from their formation, so the early phases of the ISM are recorded in
metal-poor stars. Metal-poor stars exhibit enhancements of oxygen and other
$\alpha$-elements relative to iron, which are explained by the differing yields of supernovae (SN) of type II and type Ia. The O comes from SN II (Thielemann {\it et al.} \cite{Thielemann90})), whereas Fe is produced by both SN Ia and SN II (Nomoto {\it et al.}
\cite{Nomoto84})). It is usually inferred that the
early Galaxy must have been enriched only by SN II; only more recently have SN
Ia started to pollute the ISM, reducing the $\alpha$-enhancement at higher
metallicities (Tinsley \cite{Tinsley1979}). 
This implies that the chemical composition of metal-poor halo stars is determined exclusively by SN II
events and subsequent ISM mixing in the early Galaxy.

Trends of the abundances of different elements with metallicity are not the only 
tracers of the enrichment of the Galaxy. The scatter in the relative abundances of
low-metallicity stars indicate the level of mixing of the early
ISM. Whether all of the observed scatter is an intrinsic property of the ISM, and thus
due to incomplete mixing in the early Galaxy, or is possibly due to errors that
exceed the formal estimates of uncertainties, is still unclear. 
Nucleosynthesis models invoke various sources to explain the observed abundances of heavy elements 
(e.g. Travaglio {\it et al.} \cite{Travaglio04}), rising the level of uncertainty in their nucleosynthesis production.
However, for some lighter elements, such as the $\alpha$-elements O and Mg, their expected modes of synthesis do not
predict differences in the ejected yields for stars of similar mass. Any scatter
observed in the abundances of these elements should thus reflect more directly
the effects of enrichment and mixing events in the early Galaxy.  

SN II progenitors evolve on timescales of a few to tens of Myrs. In the earliest
phase of the Galaxy there may have been a period of star formation during which the
zones enriched by SN II have not yet mixed prior to the formation of the next
stars. In this case, the newly forming stars will have different chemical
compositions. Audouze \& Silk (\cite{AS95}) proposed that, at very low
metallicity, clouds in the ISM could have been polluted by a maximum of three SN
II, consistent with the suggestion by Ryan {\it et al.} (\cite{RNB91}b) that
single SN were sufficient for the enrichment of the gas that formed stars at
[Fe/H] $\sim - 3.5$. If the early ISM was dominated by local inhomogeneities,
as predicted by inhomogeneous Galactic chemical evolution (iGCE) models (e.g. Ishimaru \& Wanajo
\cite{IW99}; Tsujimoto
\& Shigeyama \cite{Tsujimoto99}; Argast {\it et al.} \cite{Argast00}; Travaglio
{\it et al.} \cite{Travaglio01}), and the [O/Fe] and [Mg/Fe] ratios depend on
SN-progenitor mass or metallicity, then [O/Fe] and [Mg/Fe] should show a scatter
in the abundances trend versus metallicity, the surface abundances of these
stars reflecting the local enrichment of the ISM. Argast finds, for example,
that although the ISM is well mixed at [Fe/H] $> -2.0$, there is essentially no
mixing at [Fe/H] $< -3.0$. This
gives rise to progressively more star-to-star scatter as [Fe/H] decreases below
$-2.0$. 

Although there have been a large number of observational studies, the usually required agglomeration of results from different authors
introduces the possibility that the reported scatter of the $\alpha$-element
abundance may be due to a lack of internal consistency, as discussed by Norris
{\it et al.} (\cite{Norris01}). 
Underscoring this possibility, Magain (\cite{Magain87}) found that an
homogeneous re-analysis of literature data comprising 21 stars over the range
$-3.6 \leq $[Fe/H]$ \leq -0.9$ gave a mean Mg abundance [Mg/Fe]=0.45 and a
standard deviation, as a measure of the scatter, of 0.1 dex (scatters of the
order 0.15 dex were reported for [Ca/Fe] and [Si/Fe]). Ten of the
stars studied by Magain fell in the interval $-3.4 \leq $[Fe/H]$ \leq
-2.2$ which we study in the present paper. A similar result was found by Nissen
{\it et al.} (\cite{Nissen94}) for higher metallicities: [Mg/Fe]=0.41 and standard deviation 0.07 dex,
over the range $-2.0 \leq $[Fe/H]$ \leq -1.0$. Recently, Carretta {\it et al.}
(\cite{Carretta02}) found that the scatter in their estimate of intermediate
mass elemental abundances was compatible with their estimated observational
uncertainties. The most recent study of Cayrel {\it et al.} (\cite{Cayrel03}),
based on extremely high-quality VLT/UVES observations of giant stars from the HK
survey, exhibit a low scatter in the abundances of most elements,
down to metallicities as low as [Fe/H]=$-4.2$. These authors report a standard
deviation about their mean [Mg/Fe] value of 0.13 dex, and as low as 0.05 dex for
[Cr/Fe]. This indeed supports the hypothesis that there is no large scatter in
the early ISM. 

Other studies, e.g., Fuhrmann {\it et al.} (\cite{FAG95}) and Mashonkina {\it et
al.} (\cite{MGTB03}) have reported higher scatters in the alpha-element ratios.
Even though these studies suggest the existence of stars with a different
enrichment history (see e.g., Shiegeyama \& Tsujimoto (\cite{ST03}) for a
possible origin of metal-poor stars with low [$\alpha$/Fe]), their average
standard deviation is still of the order $\sim 0.1$ dex, far below what is
expected in the case of an early inhomogeneous ISM. However, among the eight
extremely metal-poor stars ([Fe/H]$<-3.5$) considered in the review by Norris
(\cite{Norris04}), two objects show [Mg/Fe]$>1.0$. This indicates the
existence of some real differences, but only for metallicity much lower than previously
thought.

Clearly, studies of the mixing efficiency in the early Galaxy are fundamental
for testing inhomogeneous GCE models. This can be accomplished
by a detailed analysis of [Mg/Fe] in halo stars, and is the main purpose of this
work. We concentrate on [Mg/Fe] because of the suitability of Mg measurements in
stars at very low metallicity, and to avoid the still debated uncertainties on
measurements of O. Studies of the scatter vs. metallicity for different stellar
populations could also allow one to constrain the yields from SN II. Attempts to
do this have so far been unsuccessful for individual stars (Chieffi \& Limongi
\cite{CL02}) because of the difficulty of computing SN yields that match the
composition of metal-poor stars. Fran\c{c}ois {\it et al.} (\cite{Francois04})
analyse how SN II yields can be constrained by the observations on the basis of
their homogeneous GCE model, suggesting that major revisions of SN II yields are
required to match the elemental abundances observed at very low metallicities.  

Iron yields are particularly difficult to predict, because Fe production depends
on the uncertainties in the mass cuts adopted for SN II models as well as the
degree of mixing prior to fallback (Nomoto {\it et al.} \cite{Nomoto97}),
whereas Mg is produced in hydrostatic burning much farther out in the star, and
hence is subject to fewer uncertainties. As a result, Mg could also be a better
chemical chronometer compared to Fe; its study would aid our understanding of the
age-metallicity relation, a key constraint on studies of chemical
evolution and nucleosynthesis.
 
The primary motivation for this study is to produce an independent estimate of
the level of elemental scatter in the early Galactic ISM, as unaffected as
possible by uncertainties. Unlike the most recent study of Cayrel {\it et al.}
(2003), whose S/N ratio is higher ( $\ge 200/1$, compared to our 100/1-150/1), and who thus
obtain better precision on a single measurement, here we focus on the
homogeneity of the sample, and demonstrate how our homogeneous chemical analysis
leads to a better constraint on the observed scatter of our derived
abundances. The aim is to avoid the introduction of star-to-star differences
that may lead to increased and misleading errors. We thus concentrate on halo
stars that are in the same evolutionary stage, have only small differences in
stellar parameters, and make use of the same Mg absorption lines for all
objects. It is of fundamental importance to understand whether the scatter we
observe is intrinsic to the sample or due to deficiencies in the analysis. We
present the results of the chemical analysis of a sample of main-sequence
turnoff halo stars, focusing on [Fe/H] and [Mg/Fe]. We describe the
definition of the sample and the observational data in \S
\ref{sample}. Particular importance has been given to the use of techniques that
produce the highest homogeneity of analysis. The methods adopted and the
chemical analysis are presented in \S \ref{analysis}. Section \ref{abundances}
reports our results and the determination of their uncertainties. In \S
\ref{discussion} we analyse the results and focus on their possible implications.

\section{Definition of the sample, observations and photometry}\label{sample}
\subsection{Data}

The stellar sample was originally selected by Ryan, Norris, \& Beers
(\cite{RNB99} -- hereafter RNB) from the surveys of Schuster \& Nissen
(\cite{Schuster88}); Ryan (\cite{Ryan89}); Beers {\it et al.}
(\cite{HK92}) and Carney {\it et al.} (\cite{Carney94}), for the purpose of
studying Li in metal-poor stars. This sample consists of 23 main-sequence
turnoff stars with a narrow range of stellar parameters, the target being
effective temperatures 6100 $\pm$ 50 K $\le$ $T_{\rm eff}$ $\le$ 6300 $\pm$ 50 K
and metallicities in the range $-3.5 \le$ [Fe/H] $\le -2.5$. The choice of
turnoff stars excludes the presence of subgiants, and reduces the range of
logarithmic surface gravity to values within a few tenths of a dex from an
expected theoretical log($g$) $\sim$ 4.0 dex. Systematic errors introduced in
the analysis will thus affect the mean values of our estimates, rather than
introducing large star-to-star differences, allowing for high internal
consistency in our derived results. 

Observations were made with the 3.9m Anglo-Australian Telescope (AAT), with the
University College London echelle spectrograph (UCLES) -- see RNB for details.
The stellar spectra are of high resolving power (R $\sim$ 40000), high
signal-to-noise ratio (S/N), exceeding 100/1 in most cases, and (incomplete)
wavelength coverage over the range $\lambda \approx$ 4900-8200 \AA . Stars were
observed in different epochs in 1996, 1997 and twice in 1998 (hereafter 1998a
and 1998b). For 12 of the 23 targets, spectra were co-added from different
epochs. At each epoch, a series of multiple observations was made to increase
the final S/N obtained. 

Photometric measurements are available from a few consistent sources, for
Johnson-Cousins $UBVRI$ colours (Ryan \cite{Ryan89}), and for Str\"omgren
$uvby$ (Schuster \& Nissen (\cite{Schuster88}; \cite{Schuster89}) and Schuster
{\it et al.} (\cite{Schuster93}; \cite{Schuster96})). For uncertainties and
interstellar reddening we adopt the estimates of RNB (and references therein)
and calculate the reddening for the $c_1$ index as $E(c_1) = 0.20 E(b-y)$ (Golay
\cite{Golay74}). As discussed below, we use photometric estimates extensively
for the stellar parameters due to the very low number of spectral lines in the
lowest-metallicity stars. The strengths of the Mg b lines at 5172 and 5183 \AA,
even at extremely low metallicity, permits Mg abundance determinations in all of
our stars.

\subsection{Galactic Populations}

The very low metallicity of most of our program stars is {\it prima facie}
evidence for their being part of the halo population. However, some recent
studies have claimed that the thick disk could extend to include stars of quite
low metallicities, [Fe/H] $\sim -2.0$, though with very low frequency. See,
for example, Allen {\it et al.} (\cite{Allen91}), Chiba \& Beers (\cite{Chiba00}), one
star of Ibukiyama \& Arimoto (\cite{IA02}), and Beers et al. (\cite{Beers02}). Therefore,
we further investigate the likely halo membership of our sample stars using the
criteria of Feltzing {\it et al.} (\cite{Feltzing03}), i.e., adopting a 200 km
s$^{-1}$ radius on a Toomre ($UW$=(U$^2$+W$^2$)$^{1/2}$ vs. V) diagram as the
border between the thick disk and halo. This limit is suggested by their data, as
well as the data of Fuhrmann (\cite{Fuhrmann98}). This is a somewhat more
detailed criteria of interpreting the actual three-dimensional motion of the
stars, as compared to criteria where only the $V$ component of motion is considered in the
identification of the population of the stars (see e.g., Ibukiyama \& Arimoto
\cite{IA02} for a description of this second method). $U$,$V$ and $W$ components were
computed by Ryan \& Norris (\cite{RN91}a) and by Carney {\it et al.}
(\cite{Carney94}) for the stars in their sample. Heliocentric velocity
components from Ryan \& Norris are transformed to the LSR frame by assuming the
solar motion relative to the LSR is (--9, 12, 7) km s$^{-1}$ in the directions
corresponding to $(U,V,W)$ (Mihalas \& Binney \cite{MB81}). The analysis in the
Toomre diagram shows that all but two of the stars in our sample lie outside the 0--200 km
s$^{-1}$ zone, i.e., they are almost certainly halo stars. Of these two exceptions,
BD$+3^{\circ}740$ has a metallicity of [Fe/H] = $-$2.78, surely a halo
metallicity, while BD$+26^{\circ}3578$ has [Fe/H] = $-$2.22, where both
estimates are based on high-S/N 1\AA\ resolution spectra (RNB). We found that
Carney's estimates of $UW$ are lower by 50 to 150 km s$^{-1}$ than those of Ryan
and Norris (\cite{RN91}a) for the six stars in common. Fuhrmann {\it et al.}
(\cite{Fuhrmann98}) and Ryan {\it et al.} (\cite{Ryan01}) found Carney {\it et
al.} photometric distances to underestimate the Hipparcos distances in each case
for nine stars in common. This possibly explains the difference in the velocity
components. A correction was applied to Carney's distances of these two stars,
which moved the velocity of BD$+3^{\circ}740$ into the halo zone, but still left
BD$+26^{\circ}3578$ in the thick-disk zone. However, our estimate of its
metallicity, [Fe/H]$=-2.49$, makes it a rather unlikely thick-disk suspect,
hence we conclude that we are analysing stars that are sampling
exclusively the halo stellar population.

\section{Chemical Analysis}\label{analysis}

In this section we describe the procedure followed to derive the abundances of
chemical elements from the observed spectra, using Kurucz's WIDTH6 code (Kurucz
\& Furenlid \cite{Kurucz79}). The code assumes local thermodynamic equilibrium
(LTE) for the formation of the spectral lines. We measure the equivalent width
(EW) for each line we consider to be reliable on the basis of a selected list of
lines that is described below. A model atmosphere with the appropriate stellar
parameters, measured EWs, and the atomic data for the spectral lines are
introduced as inputs to the code. The code then calculates the abundances for
each individual line by requiring the calculated EW to match the observed one.
The final abundance of a chemical element is the average of the line abundances.
We discuss the details of this analysis in the following sub-sections.

\subsection{Line selection and atomic data}

We have given priority to establishing the reliability of each individual
spectral line, rather than to achieving a large total number of lines. Our
initial selection is based on the lines observed in the solar spectrum (Moore
{\it et al.} \cite{Solaratlas}). We reject lines with possible blending. We do
not include lines with $gf$-values having reported uncertainties larger than 25\%, and
discard lines for which, for a large part of the sample, we derive abundances
systematically different from the mean abundance of all the lines of a specific
element. The latter can be due to cases in which the line sits at the very edge
of a spectral order, or to poorly known $gf$ data. 

Unfortunately, no single source gives laboratory $gf$ values for all lines, so
possible systematic differences between atomic data sources could introduce
systematic differences in the abundances derived from different lines. For all
lines we adopt recent experimental or solar $gf$-values and avoid theoretical
ones, because of the large discrepancies that have been found in some cases.
Whenever possible, $gf$-values are averaged from several reliable sources.
Because of our scientific aims and the paucity of lines of other species,
special attention was addressed to Fe I, Fe II and Mg I lines.

{\it Fe I -- } For Fe I lines we used mainly $gf$-values from O'Brian {\it et
al.} (\cite{OWLWB91}) and from the Oxford group (see Table \ref{gftable} for
references). However, in attempting to use Fe lines with $gf$-values listed only
by Fuhr {\it et al.} (\cite{FMW88}), we found a systematic difference in the
derived abundances of $\sim -0.1$ dex. We thus do not include lines from Fuhr
{\it et al.} alone. 
 
{\it Fe II -- } The importance of Fe II lines is to constrain the surface
gravity of the star. All but the five most metal-poor stars exhibit Fe II lines in our
spectra. We consider Fe II lines at 4923, 5197 and 5276 \AA , and
use Kroll \& Kock (\cite{KK87}) $gf$ values. Unfortunately, only for half of the
sample do we obtain good measurements; many Fe II lines are distorted by noise
or do not pass the EW threshold described below. Moreover, the Fe II line at
4923 \AA\ was identified only in the observations from 1998a, which have a
larger wavelength coverage. A comparison of the abundances derived from this
line and from the other two lines do not show any particular difference. Thus,
we believe we do not introduce any systematic error in the Fe II abundance for
the subset of stars observed in 1998a.  

{\it Mg I -- } The Mg Ib triplet around 5172 \AA\ is the most obvious feature in
all the stellar spectra of our sample. This allows us to measure Mg lines even
in the most metal-poor stars, where typically only four or five lines are usable
in the observed spectral region. We thus rely on the triplet to obtain high
star-to-star consistency in the derived Mg abundances. We also compare these
with abundances derived from the weak lines at 5528 and 5711 \AA\ whenever we
can measure them. For Mg I we adopt the solar $gf$-values of Fuhrmann {\it et
al.} (\cite{FAG95}). These are in excellent agreement with the values of Wiese
and Martin (\cite{WM80}), and the theoretical calculations of Chang
(\cite{Chang90}) and Chang \& Tang (\cite{CT90}) for the Mg Ib lines at 5172.698
and 5183.619 \AA\ and for the Mg I line 5711.092~\AA. For the line at 5528.42
\AA, we notice that the $gf$-value of --0.51 used by Fuhrmann {\it et al.} is
lower than the value --0.341 of Wiese \& Martin, but its adoption leads to
higher internal consistency of the Mg abundances derived from different lines.

The Mg Ib triplet lines are very strong, with extremely broad wings at solar
metallicities; even in our low metallicity range the line wings are still
moderately strong . Fuhrmann {\it et al.} (\cite{FAG95}) analyse Mg I lines at 4571, 4703,
4730 and the triplet at 5172 m\AA\ for 56 metal-poor stars. They report no large
systematic differences in the Mg abundances derived from these strong lines and
from weak Mg lines, nor from lines with different excitation potentials. However,
for the three stars where they identify our subset of weak lines (5528 and 5711
\AA\ ), they find abundances derived from the latter to be about 0.05 dex
smaller. In our results there is no systematic difference affecting all the
stars. Single cases in which we find a difference between weak and strong line
Mg abundances are more likely related to the uncertainties in the model
parameters (as discussed below).    

The use of the two strong Mg Ib lines requires a more careful investigation of
the damping constant. The classical treatment of the van der Waal's broadening
factor by Unsold (\cite{Unsold}) has recently been revised. The formalism of
Anstee \& O'Mara (\cite{AO}) leads to $\gamma_{\rm AO}=2.32\times\gamma_{\rm
Unsold}$, while the value derived from a fit to the Mg Ib lines in the spectrum
of the Sun is $\gamma_{\odot}=2.04\times\gamma_{Unsold}$. This suggests a $\sim$
10\% uncertainty in the damping factor, which corresponds to an uncertainty of
the order of 0.02 dex in our Mg abundances, since our Mg lines lie at the
beginning of the flat part of the curve of growth. We assume an intermediate
damping constant of $2.2\times\gamma_{Unsold}$. We note that the suggestion of
Gratton \& Sneden (\cite{GS1994}) of adopting a factor $5\times\gamma_{Unsold}$
would yield a worse agreement of the Mg abundances derived from strong and weak
lines in our sample. We do not investigate this possibility any further in this
study, and rely instead on the estimates above.
   
This selection procedure leads to a total of 52 reliable lines (including 4 Mg,
32 Fe I and 3 Fe II lines), as shown in Table~\ref{gftable}. Only eight
of these lines (of which three are Mg lines) are measurable in our most
metal-deficient stars.

\subsection{Equivalent widths}

We used IRAF to continuum fit the individual orders of the echelle spectra, and
to shift the spectra to the rest frame. Avoiding lines distorted by excessive
noise or by cosmic rays, the equivalent width (EW) of each line from our list
was measured by fitting a Gaussian profile (for lines with EWs up to $\sim$
100\AA\ ) and a Voigt profile for stronger lines (i.e., most of the Mg Ib
lines), where the Gaussian profile was inappropriate. In a few cases the adopted EW was
the average from Voigt and Gaussian fits. The Voigt fit is much more sensitive
than the Gaussian fit to the continuum location. Any systematic error in the
continuum fitting would introduce a trend in the Mg Ib abundances versus
line-strength, and thus versus metallicity (the highest-metallicity stars
showing the stronger lines). However, for these highest-metallicity stars we
always have a weaker Mg line to check the derived abundances, as discussed below. 

To estimate the uncertainties on our measured EWs, we calculated the standard
deviation for each of the 160 pairs of multiple measurements, and obtained a
mean standard deviation $\overline{\sigma}$ = 1.6 m\AA . This is in very good
agreement with the 2 m\AA\ sensitivity suggested by RNB on the basis of the
spectral S/N ratios. For the subset of stronger lines with EW $>$ 70 m\AA\ , we
found a mean standard deviation $\overline{\sigma}_{EW>70 m\AA\ }$ = 2.5 m\AA .
The strong Mg Ib lines have $\overline{\sigma}$ = 4 m\AA, corresponding to a 4\%
uncertainty in the derived EWs. 
 
\subsection{Refining the line measurements}

We filter our line measurements by constraining the line central wavelength,
measured equivalent width, and velocity width, FWHM$_{\rm v}$, corresponding to
the Gaussian FWHM. We require the centre of the fitted profile to be within 0.05
\AA\ of the central wavelength observed in the solar spectrum (Moore {\it et
al.} \cite{Solaratlas}). This allows for an uncertainty in the adopted rest-frame 
correction, and for a small deviation of the centre of the fit from the
core of the line (e.g., as can occur when the fit is dominated by the wings of the
profile). We then require a minimum EW of 7.0 m\AA, which is 4.5 times the mean
standard deviation obtained for multiple observations, $\overline{\sigma}$. An
analysis based on the more conservative criteria of requiring EW larger than
10.0 m\AA\ shows an increased scatter in our final abundances; although the
higher threshold leads to more precise measurements of individual lines, the
number of accepted lines decreases, in particular for the most metal-poor stars. The
lines were also filtered by requiring 7.0 $\leq$ FWHM$_{\rm v}$ $\leq$ 14.0 km
s$^{-1}$ in order to avoid blends, or lines so distorted by noise to appear
either narrower than the instrumental profile or unphysically wide. This selection
should lead to higher internal consistency of the measurements. The final EWs
are reported in Table \ref{gftable}. For each line we list the chemical
species (el), central wavelength ($\lambda$), adopted $gf$-value and
corresponding reference (ref), and the equivalent width measured in each program
star.

\begin{sidewaystable*}[htbp]
\caption{Equivalent widths (m\AA) for program stars. }
\label{gftable}
\centering
\begin{tiny}
\begin{tabular}{cccccccccccccc}

\hline
\hline

el&$\lambda$& log($gf$) & ref& BD--13$^{\circ}$3442& BD+1$^{\circ}$2341p& BD+20$^{\circ}$2030& BD+24$^{\circ}$1676& BD+26$^{\circ}$2621& BD+26$^{\circ}$3578& BBD+3$^{\circ}$740&BD+9$^{\circ}$2190& CD--24$^{\circ}$17504& CD--33$^{\circ}$1173\\

\hline
   Mg I & 5172.70  & --0.39    &  7	 & 113    & 109   & 118  & 134   & 119& 140& 107&   95&  79&  92\\
   Mg I & 5183.62  & --0.17    &  7	 & 128    & 122   & 155  & 148   & 134& 163& 127&  119&  92& 111\\
   Mg I & 5528.40  & --0.50    &  7	 &  20    &  20   &  26  &  32   &  18&    &  21&   14&  10&  10\\
   Mg I & 5711.10  & --1.67    &  8	 &	  &	  &	 &	 &    &    &	&     &    &	\\
   CaI  & 5581.98  & --0.56    &  12	 &	  &	  &	 &	 &    &    &	&     &    &	\\
   CaI  & 5588.76  &  0.36    &  12	 &  20    &  17   &  24  &  24   &    &  33&	&   17&    &	\\
   CaI  & 5590.13  & --0.57    &  12	 &	  &	  &	 &	 &    &    &	&     &    &	\\
   CaI  & 6122.22  & --0.31    &  13	 &  16    &  19   &  21  &  25   &    &    &  15&   12&    &	\\
   CaI  & 6162.18  & --0.09    &  13	 &  21    &  22   &  35  &  35   &  16&  42&  21&   15&    &  13\\
   CaI  & 6169.56  & --0.48    &  12	 &	  &	  &	 &	 &    &    &	&     &    &	\\
   CaI  & 6717.69  & --0.52    &  12	 &	  &	  &	 &	 &    &    &	&     &    &	\\
   Ti I & 4981.74  &   0.56    &  4,3	 &	  &  12   &  16  &  20   &    &  19&	&     &    &	\\
   Ti I & 5192.98  & --0.95    &  4	 &	  &	  &	 &	 &    &    &	&     &    &	\\
  Ti II & 5185.91  & --1.35    &  10	 &	  &	  &	 &   7   &    &    &	&     &    &	\\
   Cr I & 5206.04  &   0.02    &  10	 &	  &  17   &  25  &  21   &    &  27&	&     &    &  14\\
   Cr I & 5208.43  &   0.16    &  10	 &	  &  21   &  35  &  34   &    &  36&	&     &    &	\\
   Cr I & 5409.80  & --0.72    &  10	 &	  &	  &	 &	 &    &    &	&     &    &	\\
   Fe I & 4920.51  &   0.07    &  11	 &	  &	  &	 &	 &    &    &  31&   32&    &	\\
   Fe I & 4938.82  & --1.08    &  11	 &	  &	  &	 &	 &    &   8&	&     &    &	\\
   Fe I & 4966.10  & --0.87    &  11	 &	  &	  &	 &	 &    &    &	&     &    &	\\
   Fe I & 5041.76  & --2.20    &  11	 &	  &	  &	 &	 &    &    &	&     &    &	\\
   Fe I & 5049.83  & --1.34    &  11,1	 &	  &	  &	 &	 &    &  15&	&     &    &	\\
   Fe I & 5068.77  & --1.04    &  11	 &	  &	  &	 &   7   &    &    &	&     &    &	\\
   Fe I & 5083.35  & --2.87    &  11,2,7  &	  &	  &	 &	 &    &  10&	&     &    &	\\
   Fe I & 5166.28  & --4.20    &  5	 &	  &	  &	 &	 &    &    &	&     &    &	\\
   Fe I & 5171.61  & --1.76    &  11,6	 &  11    &  14   &  16  &  28   &  15&  26&  12&   16&    &   8\\
   Fe I & 5191.47  & --0.55    &  11	 &	  &	  &  12  &  12   &    &    &	&     &    &	\\
   Fe I & 5192.35  & --0.42    &  11	 &	  &   9   &  15  &	 &    &  21&	&     &    &	\\
   Fe I & 5194.95  & --2.06    &  11,6	 &	  &	  &   8  &  10   &    &  14&	&     &    &	\\
   Fe I & 5198.72  & --2.11    &  11,5	 &	  &	  &	 &	 &    &    &	&     &    &	\\
   Fe I & 5269.55  & --1.33    &  11	 &  57    &  54   &  66  &  73   &  61&    &  60&   54&  36&  42\\
   Fe I & 5281.80  & --0.83    &  11	 &	  &	  &	 &	 &    &  11&	&     &    &	\\
   Fe I & 5283.63  & --0.48    &  11,2	 &   7    &	  &  11  &  10   &    &  14&	&    7&    &	\\
   Fe I & 5302.31  & --0.72    &  2	 &	  &	  &	 &	 &    &    &	&     &    &	\\
   Fe I & 5324.19  & --0.10    &  2	 &	  &  13   &  20  &  24   &    &  27&  15&     &    &	\\
   Fe I & 5328.05  & --1.47    &  11,7	 &  48    &  44   &  58  &  68   &    &  70&  48&   43&    &  38\\
   Fe I & 5328.54  & --1.85    &  11	 &	  &  10   &  16  &  15   &    &  23&	&     &    &	\\
   Fe I & 5397.13  & --1.95    &  5	 &  22    &  20   &  34  &  37   &  24&    &  25&   23&  11&  14\\
   Fe I & 5405.79  & --1.86    &  11,2,7  &  27    &  22   &  34  &  40   &  28&    &  25&   23&   9&  15\\
   Fe I & 5410.92  &   0.40    &  11	 &	  &	  &	 &	 &    &   8&	&     &    &	\\
   Fe I & 5415.21  &   0.64    &  11	 &   9    &	  &  10  &  13   &    &  16&	&     &    &	\\
   Fe I & 5429.71  & --1.88    &  11	 &  28    &  20   &  35  &  44   &  32&  46&  28&   22&  11&  18\\
   Fe I & 5434.53  & --2.12    &  11,2,7  &  17    &  13   &  24  &  25   &  17&  30&  16&   13&    &  10\\
   Fe I & 5446.92  & --1.88    &  11	 &	  &  18   &  34  &  38   &  28&  41&  23&   24&  13&  11\\
   Fe I & 5569.63  & --0.49    &  1	 &	  &	  &	 &	 &    &  10&	&     &    &	\\
   Fe I & 5572.85  & --0.28    &  2	 &	  &	  &   9  &	 &    &  16&	&     &    &	\\
   Fe I & 6136.62  & --1.41    &  11	 &	  &	  &	 &	 &    &  11&	&     &    &	\\
   Fe I & 6191.57  & --1.51    &  11	 &	  &	  &	 &	 &    &   9&	&     &    &	\\
   Fe I & 6678.00  & --1.42    &  11	 &	  &	  &	 &	 &    &   7&	&     &    &	\\
  Fe II & 4923.93  & --1.32    &  9	 &	  &  38   &  46  &	 &    &    &  36&   36&    &	\\
  Fe II & 5197.58  & --2.10    &  9	 &	  &	  &	 &	 &    &  15&	&     &    &	\\
  Fe II & 5276.00  & --1.95    &  9	 &	  &	  &	 &  15   &    &    &  10&   10&    &	\\
\hline
\end{tabular}
\end{tiny}
\vspace{0.2cm}
\\
{\small 1   Bard \& Kock 1994,
2   Bard, Kock \& Kock 1991,
3   Blackwell {\it et al.}  1982a,
4   Blackwell {\it et al.} 1982b,
5   Blackwell {\it et al.} 1982c,
6   Blackwell {\it et al.} 1980,
7   Blackwell {\it et al.}  1979,
8   Fuhrmann {\it et al.}  1995,
9   Kroll \& Kock 1987,
10  Martin {\it et al.}  1988, 	  
11  O'Brian et al 1991,
12  Smith \& Raggett 1981,
13  Wiese \& Martin 1980. }
\end{sidewaystable*}

\begin{sidewaystable*}[htbp]
\centering
\begin{tiny}
\begin{tabular}{ccccccccccccccc}
\hline
\hline

el&$\lambda$& CD--35$^{\circ}$14894& CD--71$^{\circ}$1234& CD22943--045& G126--52& G4--37&G64--12&G64--37& HD74000&HD84937& LP635--14& LP651----4& LP815--43 &LP831--70 \\
\hline
   Mg I & 5172.70 & 154& 135& 151& 146& 140&   77 &  77& 192& 160& 132& 125&  95& 113\\
   Mg I & 5183.62 & 179& 152& 177& 157& 166&   92 &  93& 234& 180& 154& 146& 118& 138\\
   Mg I & 5528.40 &    &  30&  45&    &    &	8 &  10&    &  47&  33&  26&	&  21\\
   Mg I & 5711.10 &    &    &	 &    &    &	  &    &   9&	 &    &    &	&    \\
   CaI & 5581.98 &    &    &  10&    &    &	  &    &    &	 &    &    &	&    \\
   CaI & 5588.76 &    &  25&  36&  32&  22&	  &    &  41&  36&  24&  14&	&    \\
   CaI & 5590.13 &    &    &	8&    &    &	  &    &  11&	 &    &    &	&    \\
   CaI & 6122.22 &    &  21&	 &    &    &	  &    &    &  35&  19&    &	&    \\
   CaI & 6162.18 &  36&  34&  43&  38&    &   10 &   9&  57&  46&  31&  23&  17&    \\
   CaI & 6169.56 &    &    &  10&    &    &	  &    &  13&	 &    &    &	&    \\
   CaI & 6717.69 &    &    &	 &    &    &	  &    &    &	8&    &    &	&    \\
   Ti I & 4981.74 &    &    &  20&  18&  14&	  &    &  29&  23&    &    &	&  11\\
   Ti I & 5192.98 &    &    &	 &    &    &	  &    &   8&	 &    &    &	&    \\
  Ti II & 5185.91 &    &    &	8&    &    &	  &    &    &  10&    &    &	&    \\
   Cr I & 5206.04 &  27&  22&  29&  31&  27&	  &    &  45&  35&  22&    &	&    \\
   Cr I & 5208.43 &    &  34&  40&    &  30&	  &    &  65&	 &  32&    &	&  15\\
   Cr I & 5409.80 &   7&  14&	 &    &    &	  &    &  14&	9&    &    &	&    \\
   Fe I & 4920.51 &    &    &	 &    &    &	  &    &    &	 &    &    &	&    \\
   Fe I & 4938.82 &   8&   7&	7&   8&    &	  &    &  19&	8&    &    &	&    \\
   Fe I & 4966.10 &    &    &	 &    &    &	  &    &  15&	 &    &    &	&    \\
   Fe I & 5041.76 &    &  11&	 &    &    &	  &    &    &	 &    &    &	&    \\
   Fe I & 5049.83 &  15&  12&  14&  16&  13&	  &    &  30&  19&  15&   9&	&    \\
   Fe I & 5068.77 &   9&    &	 &   8&    &	  &    &  18&	 &    &    &	&    \\
   Fe I & 5083.35 &  11&    &	 &  10&    &	  &    &  17&	9&   8&    &	&    \\
   Fe I & 5166.28 &    &    &	 &    &    &	  &    &   8&	 &    &    &	&    \\
   Fe I & 5171.61 &  29&  23&  25&    &  23&	  &   8&  42&  32&  20&  14&  11&  11\\
   Fe I & 5191.47 &  14&  15&  17&  18&    &	  &    &  31&  22&   9&    &	&   9\\
   Fe I & 5192.35 &  21&  18&  19&  24&  14&	  &    &  39&	 &  15&    &	&    \\
   Fe I & 5194.95 &  16&  13&  14&  14&    &	  &    &  25&  19&  10&  14&	&    \\
   Fe I & 5198.72 &    &    &	 &    &    &	  &    &   8&	 &    &    &	&    \\
   Fe I & 5269.55 &    &  73&	 &    &    &   32 &  34&    &	 &  70&  56&	&  52\\
   Fe I & 5281.80 &  10&    &  13&    &   8&	  &    &  20&  11&   7&    &	&    \\
   Fe I & 5283.63 &  15&  11&  17&  15&   9&	  &    &  29&  18&  12&   9&	&    \\
   Fe I & 5302.31 &  10&    &	8&   9&    &	  &    &  16&	8&    &    &	&    \\
   Fe I & 5324.19 &  26&  22&  27&  25&  22&	  &    &  42&  32&  20&  15&	&    \\
   Fe I & 5328.05 &    &  62&  68&  71&  67&	  &  31&  88&  77&  64&  47&  38&  41\\
   Fe I & 5328.54 &    &  21&  23&  25&  16&	  &    &  41&  29&  17&   8&	&    \\
   Fe I & 5397.13 &    &  37&  43&    &    &	9 &  12&    &	 &  39&  27&	&  23\\
   Fe I & 5405.79 &  49&  42&  47&    &  38&   13 &  15&  68&  52&  38&  28&  21&  19\\
   Fe I & 5410.92 &    &  10&	8&   8&    &	  &    &  16&  10&    &    &	&    \\
   Fe I & 5415.21 &  17&  13&  15&  16&    &	  &    &  29&  19&  12&    &	&    \\
   Fe I & 5429.71 &  48&  44&  46&  48&  39&	9 &  13&  69&  55&  39&  31&  21&  24\\
   Fe I & 5434.53 &  33&  27&  31&  35&  25&	  &   8&  51&  39&  26&  18&  13&  14\\
   Fe I & 5446.92 &    &  40&  44&  47&  37&	  &  13&  65&  51&  37&  27&  21&  21\\
   Fe I & 5569.63 &  10&   9&	9&   9&    &	  &    &  20&  15&   8&    &	&    \\
   Fe I & 5572.85 &  16&  11&  16&  14&    &	  &    &  31&  23&  11&    &   7&    \\
   Fe I & 6136.62 &  13&   8&	 &  10&    &	  &    &  25&  16&  10&   9&	&    \\
   Fe I & 6191.57 &    &    &	 &    &    &	  &    &  20&  13&   9&    &	&    \\
   Fe I & 6677.80 &   8&    &  11&    &    &	  &    &  17&  11&   7&    &	&    \\
  Fe II & 4923.93 &    &  48&	 &    &    &	  &  22&    &  63&    &    &	&    \\
  Fe II & 5197.58 &   9&   8&  14&  12&    &	  &    &  15&  15&    &    &	&    \\
  Fe II & 5276.00 &  12&  12&  15&    &    &	  &    &    &  20&  10&    &	&    \\
\hline
\end{tabular}
\end{tiny}
\end{sidewaystable*}

\subsection{Model Atmosphere}


We adopt the grid of 1D, LTE model atmospheres of Kurucz (1989, private
communication), interpolating within the original models to our required stellar
parameters. As previously noted (Ryan {\it et al.} \cite{RNB96}), abundances
derived with Kurucz's 1989 models are in good agreement with abundances derived
with the model atmospheres of Bell {\it et al.} (\cite{Bell81}), but newer
Kurucz (\cite{Kurucz93}) models have a different temperature structure and lead
to abundances higher than the previous models. Ryan {\it et al.} (\cite{RNB96})
found that temperatures are up to 200 K higher in the line-forming region of the
Kurucz (\cite{Kurucz93}) models and result in +0.1 dex in their Fe I abundances
for turn-off stars. We find a +0.1 dex difference for lines of EW $\sim$ 10
m\AA, rising to +0.15 dex at $\sim$ 40 m\AA\ and to +0.2 dex at $\sim$ 150 m\AA,
the wings of stronger lines being more sensitive to these regions. The two
strong Mg Ib lines are thus more sensitive to the change in the temperature
structure of the model than the bulk of Fe lines. This would imply a typical
--0.05 dex difference in the [Mg/Fe] ratio if we derived the abundances with the
Kurucz (\cite{Kurucz93}) models and only the b lines of Mg I. However, the use
of weak Mg lines allows us also to use estimates for the abundance from lines in
the same line-forming region as the Fe lines, and thus obtain a [Mg/Fe] value
that is less dependent on the temperature structure of the model. The average of
Mg abundances from weak and strong lines reduces the dependence of our results
on the adopted model. Moreover, for the lowest-metallicity stars (where weak Mg
lines are not measurable), the Mg Ib lines are also weaker and thus less
dependent on the temperature structure. We estimate that if we were to adopt
Kurucz (\cite{Kurucz93}) models, the standard deviation about the mean [Mg/Fe]
would increase by less than 0.005 dex. We can thus safely neglect the model
dependence in our discussion of the observed scatter.

\subsection{Stellar Parameters}\label{stellarparameters}

   \begin{figure}
   \centering
   \includegraphics[width=88mm]{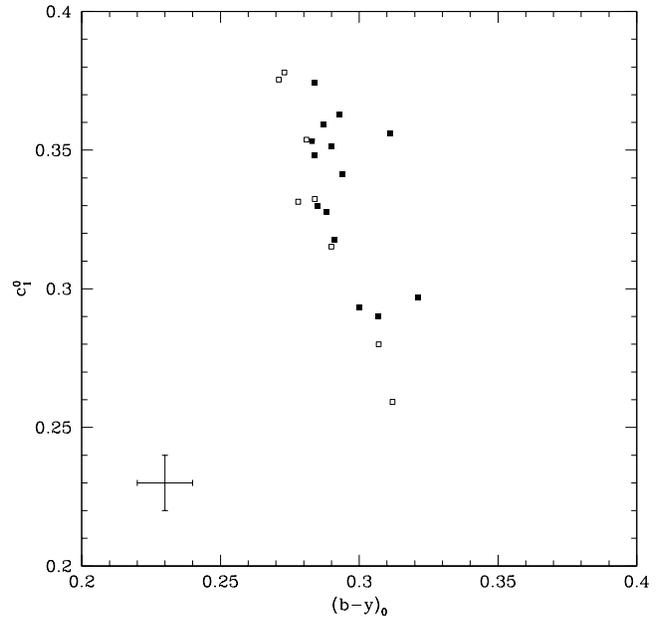}
      \caption{ The position in the diagram dereddened c$_1$ vs. $b-y$ is shown as proof of the similar evolutionary stage of the stars. Filled squares represent stars with measurements of Fe II lines and thus spectroscopic estimate of log($g$). Empty squares are used for stars with no Fe II measurements.   
              }
         \label{photo}
   \end{figure}
%

In principle, we can derive the effective temperature, $T_{\rm eff}$,
in an iterative fashion by computing abundances for several models, and requiring no
dependence of the Fe abundance on excitation potential (EP). The surface gravity
log($g$) can be set by requiring the Fe abundance derived from Fe II lines to
match that from Fe I lines. Finally, the microturbulent velocity can be derived
by requiring no trend of the abundances versus EW. However, these spectroscopic
estimates for the stellar parameters rely on the measurement of at least {\it
some} lines of {\it both} Fe I and Fe II. The higher the number of lines, the higher the
precision of the estimate. As discussed before, the very metal-poor stars show
just four or five reliable lines in the stellar spectra; for about half of
the sample we cannot measure any Fe II lines at all. Hence, we are not able to
derive spectroscopic estimates of the parameters for all the stars and furthermore,
when we do have estimates, not all have the same precision. 

Photometry can assist us in the task of determining the stellar parameters,
having the excellent characteristic of not biasing our precision at lowest
metallicity (as is the case for spectroscopic estimates, which rely on a
decreasing number of lines at lowest metallicity). As shown in Fig.
\ref{photo}, the stars are in a region of the $c_1$ vs. $b-y$ diagram that led
RNB to identify them as main-sequence turnoff stars. This determination is done
by comparing the observations with evolutionary tracks, on the basis of the
$b-y$ temperature sensitive index and the gravity sensitivity of the $c_1$
index. We go a step further, and obtain estimates for the surface gravity by
calibrating the $c_1$ index with stars whose spectroscopic gravities we are
able to derive. Because of the small number of Fe I lines and the absence of Fe II lines
for the lowest-metallicity stars, we adopt photometric calibrations both for
$T_{\rm eff}$ and log($g$), based on the homogeneity of the sample, rather than
using solely spectroscopic values. We discuss the details of our method below.

{\it Effective Temperature -- }

The effective temperature is adopted from RNB. Their photometric estimates were
based on six indices (B$-$V, V$-$R, R$-$I, $\beta$, HP2 and $b-y$), calibrated
with the theoretical calculations of Bell \& Oke (\cite{BO86}) and the empirical
relations of Magain (\cite{Magain87}). The photometric indices were de-reddened
using the reddening maps of Lucke \cite{Lucke78} and Burstein \& Heiles
\cite{BH82}), with Johnson photometric distances from Carney {\it et al.} (\cite{Carney94}) and
Ryan {\it et al.} (\cite{Ryan89}), and the comparison of $b-y$ with the
reddening-free index $\beta$ for Str\"omgren indices. A correction was applied
to account for a 0.020 mag difference between the colour excess estimate derived
from Str\"omgren indices and that inferred from reddening maps, where we expect
E($b-y$)=0.7E(B$-$V). The uncertainties on the photometric indices were assumed
to be on average less than $\pm$0.01 magnitudes (see RNB and references therein
for details). The final $T_{\rm eff}$ estimates are shown in Table
\ref{tableparameters}. They exhibit a very low relative uncertainty, on the of
order $\pm$40 K, although larger systematic errors could affect the zero point
by up to $\sim$ 100 to 200 K.

Later spectroscopic reanalysis of some stars in our sample have suggested higher
temperatures (see e.g., Nissen {\it et al.} \cite{Nissen01}; Ford {\it et al.}
\cite{Ford02}). We note the problem of systematic differences between
photometric and spectroscopic estimates of $T_{\rm eff}$, the former being
typically expected to be 100 K hotter according to Alonso {\it et al.}
(\cite{Alonso99}). Johnson (\cite{Johnson02}) finds up to 150K difference
between $T_{\rm eff}$ estimates derived with spectroscopy or with photometry. We
therefore determine a spectroscopic estimate of the $T_{\rm eff}$ for the stars
by requiring no dependence of the abundance derived from Fe I lines versus the
excitation potential (EP) of the line. For most of the stars, there are too few
Fe I lines to have a high confidence in the determination of spectroscopic
$T_{\rm spec}$; we work on the subset of 15 stars with more than 10 lines. Our
estimates for the effective temperature $T_{\rm spec}$ are on average 80K lower
than the temperatures derived by RNB. This is just at the limit of our average
1$\sigma$ statistical error of 80K in the derivation of $T_{\rm spec}$. However,
in most cases we have only two or three Fe I lines with high EP, hence the
procedure is heavily dependent on these few lines. We thus rely instead on the
photometric estimates from RNB, and assess the impact of a possible systematic
error up to 150K, as suggested by the largest differences between RNB estimates
and our spectroscopic determinations. We assume a 40K random error on the
temperature from RNB.
 
\begin{table}[htbp]                                    
\centering
\caption{Model parameters}\label{tableparameters}
\begin{tabular}{l|ccccc}
\hline
\hline

star	 &   $T_{\rm RNB}$ &$\sigma_{T}$ &$\xi $ &log($g$)$_{\rm spec}$ &log$(g)_{\rm phot}$  \\
&&&$(\pm 0.2)$&&$(\pm 0.2)$\\
\hline

BD-13$^{\circ}$3442   &6210   &30     &1.5    &--     &3.6\\
BD+1$^{\circ}$2341p   &6260   &40     &1.5    &3.8    &3.8\\
BD+20$^{\circ}$2030   &6200   &40     &1.6    &3.9    &3.9\\
BD+24$^{\circ}$1676   &6170   &30     &1.8    &3.7    &3.7\\
BD+26$^{\circ}$2621   &6150   &40     &1.6    &--     &4\\
BD+26$^{\circ}$3578   &6150   &40     &1.5    &3.4    &3.7\\
BD+3$^{\circ}$740     &6240   &40     &1.8    &3.7    &3.7\\
BD+9$^{\circ}$2190    &6250   &30     &1.3    &3.8    &3.6\\
CD--24$^{\circ}$17504 &6070   &30     &1.8    &--     &4.2\\
CD--33$^{\circ}$1173  &6250   &20     &1.8    &--     &3.7\\
CD--35$^{\circ}$14894 &6060   &20     &1.3    &4.0    &4.1\\
CD--71$^{\circ}$1234  &6190   &30     &1.5    &4.1    &4.1\\
CS22943--045	      &6140   &40     &1.5    &3.7    &3.9\\
G126--52	      &6210   &40     &1.8    &3.9    &3.8\\
G4--37		      &6050   &40     &1.8    &--     &4.1\\
G64--12 	      &6220   &30     &1.7    &--     &3.9\\
G64--37 	      &6240   &30     &1      &3.9    &3.9\\
HD 74000	      &6040   &30     &1.3    &4.2    &4.1\\
HD84937 	      &6160   &30     &1.8    &3.9    &3.7\\
LP635--14	      &6270   &30     &1.7    &4.4    &3.7\\
LP651--4	      &6240   &30     &1      &--     &3.9\\
LP815--43	      &6340   &30     &1      &--     &3.6\\
LP831--70	      &6050   &20     &1      &--     &4.3\\

\hline
\end{tabular}
\end{table}

{\it Surface gravity and microturbulent velocity -- }

We adopt an iterative procedure to determine both the surface gravity and the
microturbulent velocity of each star. We make calculations with log($g$)=4.0 and
$\xi$=1.0--1.5 km s$^{-1}$ as our first stage. With these results we calculate
the best estimate for $\xi$ by requiring no trend of the abundances of Fe I with
the EW of the line the abundance was derived from. Due to the lack of lines for
the most metal-poor stars, we adopt a range for $\xi$ of 1.0--1.5 s$^{-1}$ to
compensate for the poor spectroscopic determination. We perform calculations
with the assumed log($g$)$\pm$0.2 dex and the calculated $\xi$, searching for the
value of log($g$) that would let the Fe abundances derived from Fe I match the
ones from Fe II lines. We then iterate this procedure to obtain new estimates of
$\xi$ and log($g$), finding a convergence of the results after a few iterations. 

Only one or two Fe II lines are available for each log($g$) determination. The
standard error in the difference [Fe/H]$_{\rm I}$--[Fe/H]$_{\rm II}$ corresponds
to an error $\sigma_{{\rm log}(g)_{\rm Fe}}\sim$ 0.12 dex. The inferred log($g$) value
is also affected by the adopted effective temperature. Changes in $T_{\rm eff}$
of 40 K cause a variation in the inferred log($g$) of $\sigma_{ {\rm log}(g)_T}\sim$
0.02 dex. Values generally adopted for $\xi$ in the case of turn-off stars are
in the range 1.0 -- 1.5 km s$^{-1}$. A conservative assumption of a variation of
$\xi$, over the range $\Delta\xi$=$\pm$ 0.2 km s$^{-1}$, produces a variation in
the derived log($g$) of $\pm$ 0.06 dex on average. On the other hand, a variation
of log($g$) of $\pm$ 0.06 produces a variation in $\xi$ of less than 0.03 km
s$^{-1}$. Thus the determination of $\xi$ is quite insensitive to the adopted
log($g$). We expect this 0.06 dex uncertainty on log($g$) to arise only in the worse
cases when we cannot constrain $\xi$; in most cases this iterative procedure
produces an uncertainty in the derived log($g$) of $\sigma_{ {\rm log}(g)_{\rm procedure}}
\sim$ 0.04 dex. We thus assume a random error on the spectroscopic log($g$) of
$\sigma_{{\rm log}(g)_{\rm spec}}=\sqrt{\sigma_{ {\rm log}(g)_{\rm Fe}}^2+\sigma_{ {\rm log}(g)
_T}^2+\sigma_{{\rm log}(g)_{\rm procedure}}^2}=0.13$. For $\xi$ we adopt the more
conservative uncertainty of 0.2 km s$^{-1}$.

   \begin{figure}
   \centering
   \includegraphics[width=88mm]{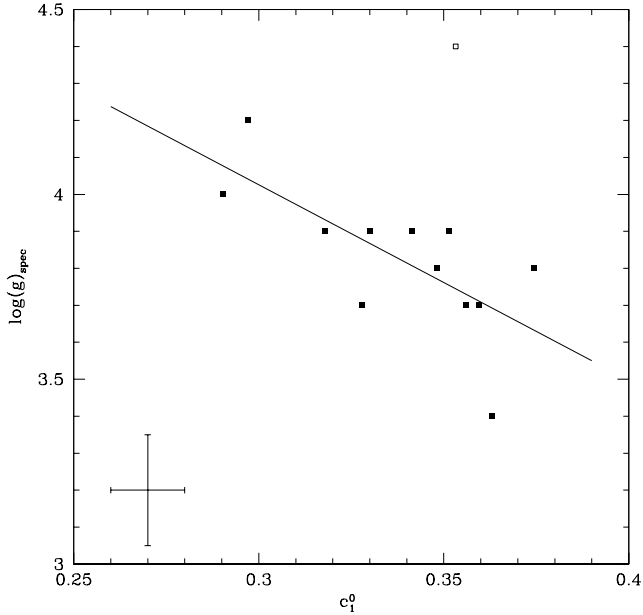}
      \caption{ Calibration of the spectroscopic log($g$) vs the gravity sensitive dereddened c$_1$ photometric index. The empty square represents the star LP635-14 that was rejected from the fit by a 2$\sigma$ clipping procedure.   
              }
         \label{gcalibration}
   \end{figure}
%

These spectroscopic estimates of log($g$) are used to calibrate log($g$) versus the
de-reddened $c_{1}$ index using a linear fit (see Fig. \ref{gcalibration}). The
c$_1$ index was de-reddened adopting the colour excess law $E(c_1)=0.20E(B-V)$
and $c_{1,0}=c_{1}-E(c_{1})$ (Golay \cite{Golay74}). The calibration is then
used to calculate  'photometric' estimates of the log($g$) for each star (Table
\ref{tableparameters}).  The standard deviation about the fit for the
calibration of log($g$) versus $c_1$ is 0.15 dex, excluding (by a 2$\sigma$
clipping procedure) the log($g$)=4.4 derived spectroscopically from just one Fe II
line for LP635--14. The error on the photometric estimates of log($g$) is adopted
to be 0.2 dex, as the quadratic sum of the uncertainty on the spectroscopic
log($g$) (0.13) and of the calibration versus $c_1$ (0.15), the influence of the
uncertainty on the $c_1$ value being negligible. We then derive the abundances
with the adopted photometric log($g$) for the entire sample of stars. For the
subset of stars for which we could calculate the spectroscopic log($g$), we
perform a further analysis with that value of log($g$), but find no significant
difference in the overall trend or scatter of the results.  

We investigate a possible dependence of the c$_1$ index on temperature as well
as gravity, but find no evidence of a correlation between the residuals
of the gravity calibration and the temperature scale. This indicates that the
gravity sensitivity of the c$_1$ index is far more important than its
temperature sensitivity over the narrow temperature range of our stars, and the
latter does not affect our calibration. 

{\it Initial chemical composition -- }

We also explored the possible sensitivity of the derived abundances on the initial
chemical composition adopted for the calculations. The chemical composition is
introduced in the model atmosphere by scaling the solar composition with the
metallicity of the star. However, because of the very low metallicity of our
stars, no significant change in the derived abundances is produced by adopting
models with slightly different chemical composition.

\section{Chemical abundances} \label{abundances}


We now derive surface abundances from lines of Mg I and Fe I for our entire
sample of stars. For most of the stars we are also able to derive surface
abundances for the elements Ca, Ti, and Cr, depending on the presence of
reliable spectral lines in our wavelength coverage. The results are listed in
Tables \ref{tableabundances} and \ref{tableothers}, along with the number of
lines used, as well as the standard deviation, $\sigma$, of the abundances derived
from different lines. The total uncertainty of each element is described below.

\begin{table*}[htbp]                                    
\centering
\caption{Fe and Mg chemical abundances.}\label{tableabundances}
\vspace{2mm}

\begin{tabular}{l|cccccccccc}
\hline
\hline

star	&  [Fe/H]$_{\rm I}$ 	&n$_{\rm Fe I}$  	&$\sigma_{\rm Fe I}$ 
&  [Fe/H]$_{\rm II}$	&n$_{\rm Fe II}$ 	&$\sigma_{\rm Fe II}$ 
&  [Mg/H]			&n$_{Mg}$		&$\sigma_{\rm Mg}$&[Mg/Fe]\\

\hline
		     
BD-13$^{\circ}$3442   &--2.83  &9      &0.02    &---    &0     &---    &--2.47 &3&0.03	&0.36\\
BD+1$^{\circ}$2341p   &--2.88  &11     &0.02    &--2.85  &1     &---    &--2.51 &3&0.01	&0.37\\
BD+20$^{\circ}$2030   &--2.68  &16     &0.01    &--2.68  &1     &---    &--2.39 &3&0.06	&0.29\\
BD+24$^{\circ}$1676   &--2.60  &15     &0.02    &--2.60  &1     &---    &--2.28 &3&0.01	&0.32\\
BD+26$^{\circ}$2621   &--2.86  &7      &0.01    &---    &0     &---    &--2.59 &3&0.01	&0.27\\
BD+26$^{\circ}$3578   &--2.49  &21     &0.01    &--2.40  &1     &---    &--2.16 &2&0.01	&0.33\\
BD+3$^{\circ}$740     &--2.84  &10     &0.01    &--2.87  &2     &0.09   &--2.52 &3&0.03	&0.32\\
BD+9$^{\circ}$2190    &--2.83  &10     &0.02    &--2.87  &2     &0.07   &--2.61 &3&0.07	&0.22\\
CD--24$^{\circ}$17504 &--3.45  &5      &0.03    &---    &0     &---    &--3.17 &3&0.10	&0.28\\
CD--33$^{\circ}$1173  &--3.10  &8      &0.03    &---    &0     &---    &--2.81 &3&0.04	&0.29\\
CD--35$^{\circ}$14894 &--2.55  &20     &0.01    &--2.53  &2     &---    &--2.31 &2&0.01	&0.24\\
CD--71$^{\circ}$1234  &--2.55  &22     &0.01    &--2.56  &3     &0.01   &--2.35 &3&0.03	&0.20\\
CS22943--045	      &--2.50  &22     &0.02    &--2.43  &2     &0.08   &--2.13 &3&0.02	&0.37\\
G126--52	      &--2.44  &20     &0.01    &--2.49  &1     &---    &--2.17 &2&0.05	&0.27\\
G4--37		      &--2.75  &12     &0.02    &---    &0     &---    &--2.48 &2&0.00	&0.27\\
G64--12 	      &--3.35  &4      &0.04    &---    &0     &---    &--3.08 &3&0.05	&0.27\\
G64--37 	      &--3.15  &8      &0.02    &--3.13  &1     &---    &--2.92 &3&0.04	&0.23\\
HD 74000	      &--2.17  &28     &0.01    &--2.24  &1     &---    &--1.95 &3$^1$&0.07&0.22\\
HD84937 	      &--2.36  &23     &0.01    &--2.44  &3     &0.01   &--2.01 &3&0.02	&0.35\\
LP635--14	      &--2.53  &23     &0.02    &--2.80  &1     &---    &--2.19 &3&0.02	&0.34\\
LP651--4	      &--2.69  &14     &0.03    &---    &0     &0.04   &--2.27 &3&0.04	&0.42\\
LP815--43	      &--2.81  &7      &0.01    &---    &0     &---    &--2.42 &2&0.09	&0.39\\
LP831--70	      &--3.06  &9      &0.03    &---    &0     &---    &--2.69 &3&0.06	&0.37\\

\hline
\end{tabular}
\\
\vspace{0.2cm}
{\small $^1$: In HD74000 we measured the Mg weak line at 5711 m\AA\ instead of the line at 5528 m\AA. }
\end{table*}

\begin{table*}[htbp]                                    
\centering
\caption{Ca, Ti and Cr chemical abundances. Ti II abundances are derived only in 3 stars.}\label{tableothers}
\vspace{2mm}

\begin{tabular}{l|ccccccccc}
\hline
\hline

star	&  [Ca/H] 	&n$_{\rm Ca}$  	&$\sigma_{\rm Ca}$ 
	&  [Ti/H]	&n$_{\rm Ti}$ 	&$\sigma_{\rm Ti}$ 
	&  [Cr/H]	&n$_{\rm Cr}$	&$\sigma_{\rm Cr}$\\
\hline
		     
BD-13$^{\circ}$3442  	 &--2.4   &3	 &0.03	 &---		 &0	 &---	 &---	  &0  &--- \\
BD+1$^{\circ}$2341p  	 &--2.37  &3	 &0.04	 &--2.25 	 &1	 &---	 &--2.84  &2  &0.01\\
BD+20$^{\circ}$2030  	 &--2.23  &3	 &0.04	 &--2.16 	 &1	 &---	 &--2.62  &2  &0.04\\
BD+24$^{\circ}$1676  	 &--2.21  &3	 &0.03	 &--2.07$^1$	 &1	 &---	 &--2.72  &2  &0.08\\
BD+26$^{\circ}$2621  	 &--2.67  &1	 &---	 &---		 &0	 &---	 &---	  &0  &--- \\
BD+26$^{\circ}$3578  	 &--2.05  &2	 &0.02	 &--2.10 	 &1	 &---	 &--2.62  &2  &0.04\\
BD+3$^{\circ}$740    	 &--2.43  &2	 &0.01	 &---		 &0	 &---	 &---	  &0  &--- \\
BD+9$^{\circ}$2190   	 &--2.51  &3	 &0.06	 &---		 &0	 &---	 &---	  &0  &--- \\
CD--24$^{\circ}$17504	 &---	 &0	 &---	 &---		 &0	 &---	 &---	  &0  &--- \\
CD--33$^{\circ}$1173 	 &--2.69  &1	 &---	 &---		 &0	 &---	 &--2.96  &1  &0.0 \\
CD--35$^{\circ}$14894	 &--2.22  &1	 &---	 &---		 &0	 &---	 &--2.69  &2  &0.07\\
CD--71$^{\circ}$1234 	 &--2.23  &3	 &0.02	 &---		 &0	 &---	 &--2.51  &3  &0.18\\
CS22943--045	     	 &--1.99  &5	 &0.05	 &--2.08$^2$	 &1	 &---	 &--2.57  &2  &0.04\\
G126--52	     	 &--2.09  &2	 &0.00	 &--2.07 	 &1	 &---	 &--2.54  &1  &--- \\
G4--37		     	 &--2.39  &1	 &---	 &--2.36 	 &1	 &---	 &--2.81  &2  &0.04\\
G64--12 	     	 &--2.87  &1	 &---	 &---		 &0	 &---	 &---	  &0  &--- \\
G64--37 	     	 &--2.86  &1	 &---	 &---		 &0	 &---	 &---	  &0  &--- \\
HD 74000	     	 &--1.93  &4	 &0.05	 &--1.91 	 &2	 &0.03	 &--2.24  &3  &0.08\\
HD84937 	     	 &--1.94  &4	 &0.04	 &--1.97$^3$	 &1	 &---	 &--2.44  &2  &0.04\\
LP635--14	     	 &--2.22  &3	 &0.02	 &--1.78 	 &1	 &---	 &--2.63  &2  &0.04\\
LP651--4	     	 &--2.43  &2	 &0.07	 &---		 &0	 &---	 &---	  &0  &--- \\
LP815--43	     	 &--2.47  &1	 &---	 &---		 &0	 &---	 &---	  &0  &--- \\
LP831--70	     	 &---	 &0	 &---	 &--2.46 	 &1	 &---	 &--3.20  &1  &--- \\

\hline
\end{tabular}
\\
\vspace{0.2cm}
{\small $^1$: [Ti/H]$_{\rm II}=-2.19$,
$^2$: [Ti/H]$_{\rm II}=-2.09$ and 
$^3$: [Ti/H]$_{\rm II}=-2.03$.}
\end{table*}

\subsection{Dependence of abundances on model atmosphere parameters}\label{parametersdependence}

The dependence of our derived abundances on stellar parameters varies with the
strength of the lines used. Because of the homogeneity of our sample, we study this
dependence for the whole sample rather than for single stars. 

For Fe I, the number of weak lines is far higher than the number of stronger
ones. This reduces the sensitivity to strong lines, hence there will be no large
differences of sensitivity between higher and lower metallicity stars. Fe II
lines are never stronger than 60 m\AA\ ,  so again the averaged Fe II sensitivity
is a good representative for all the stars. We derive abundances for a grid of
models in which we vary independently one stellar parameter at a time by its
standard error, as discussed in Sec. \ref{stellarparameters}, and report the
average results for Fe lines in Table \ref{tabledependence}. 

In the case of Mg, the model parameter sensitivity is much more significant in
the Mg Ib lines than in the weak Mg line. Besides a dependence on temperature
similar to that of Fe I lines, Mg Ib lines show an extremely high sensitivity to
surface gravity because of the stronger wings. We show in Table
\ref{tabledependence} the average results for the weak Mg I line at 5528 \AA\
and for the average Mg Ib lines. The dependence of the [Mg/Fe] ratio is averaged
on the whole sample of stars, and thus weighted on both the weak line and the two
strong lines.

\begin{table*}[htbp]                                    
\centering
\caption{Dependence of abundances on model parameters}\label{tabledependence}
\vspace{2mm}
\begin{tabular}{lcccccc}
\hline
\hline
\vspace{0.2cm}
parameter&variation&$\overline{\Delta{\rm [Fe/H]}_{\rm I}}$&$\overline{\Delta{\rm [Fe/H]}_{\rm II}}$&$\overline{\Delta[{\rm Mg}/{\rm H}]_{\rm 5528}}$&$\overline{\Delta[{\rm Mg}/{\rm H}]_{\rm Ib}}$&$\overline{\Delta[{\rm Mg}/{\rm Fe}]}$\\
\hline
$T_{\rm eff}$ (\it random)& $\pm 40 {\rm K}$  & $\pm 0.03$ & $\pm 0.00$ & $\pm 0.02$ & $\pm 0.04$ & $\pm 0.00$ \\
$T_{\rm eff}$ (\it systematic) & $\pm 150 {\rm K}$  & $\pm 0.13$ & $\pm 0.02$ & $\pm 0.07$ & $\pm 0.14$ & $\pm 0.01$ \\
$\xi$     & $\pm 0.2$ & $\mp 0.01$ & $\pm 0.00$ & $\mp 0.01$ & $\mp 0.03$ & $\mp 0.02$ \\
log($g$)    & $\pm 0.2$  & $\mp 0.01$ & $\pm 0.03$ & $\mp 0.02$ & $\mp 0.10$ & $\mp 0.06$ \\
\hline
\end{tabular}
\end{table*}

Estimating uncertainties by varying just one parameter at a time is a useful
exercise, but it does not reflect the real analysis procedure. In order to estimate more
realistic uncertainties, we performed a Monte Carlo simulation reproducing the
procedure adopted for deriving log($g$) and $\xi$. We consider two candidate
stars, BD$+3^{\circ}740$ (10 Fe I lines, [Fe/H] = --2.84) and
BD$+26^{\circ}3578$ (21 Fe I lines, [Fe/H] = --2.49) as representative of our
sample. For each star we create a set of 25 virtual "observations" by adding
numerical noise to the measured EWs. For each line, the distribution of the 25
randomly generated errors added to the EW is a Gaussian with $\sigma$ = 1.5
m\AA\ (i.e., our average inferred 1$\sigma$ uncertainty on the EW). By deriving
the surface gravity and microturbulent velocity for the virtual observations
with the iterative method used for the real data, we were able to estimate the
dependence of our abundances {\it and} derived parameters on observational
errors, by means of the standard deviation of the results for the 25 simulations. In
Table \ref{tablerandom} we report the results of this ``end-to-end'' test of our
procedures.

\begin{table*}[htbp]                                    
\centering
\caption{Standard deviations for the results of the Monte Carlo test} 
\vspace{2mm}

\label{tablerandom}
\begin{tabular}{lccccccccc}
\hline
\hline

star&[Fe/H] & $\sigma_{\rm EW}/$m\AA\ & $\sigma_{\xi}$ & $\sigma_{\rm log}(g)$ & {$\sigma_{\rm [Fe/H]_{\rm I}}$} & {$\sigma_{\rm [Fe/H]_{\rm II}}$} & $\sigma_{\rm [Mg/H]}$ & $\sigma_{\rm [Mg/Fe]}$ & $\sigma_{\rm [Ca/Fe]}$\\
\hline
BD$+3^{\circ}740$&--2.84&1.5&0.20&0.19&0.02&0.03&0.06&0.06&0.02\\
BD$+26^{\circ}3578$&--2.49&1.5&0.17&0.19&0.02&0.03&0.04&0.03&0.02\\
\hline
\end{tabular}
\end{table*}         

From this test we find that the sensitivity of our results is {\it lower} than a
quadratic sum of the single parameter dependences estimated one at a time. This
is due to the fact that the errors in the parameters derived with the iterative
procedure are not independent, so the parameters partially compensate for one
another. Moreover, the abundances are averaged on both weak and stronger lines,
decreasing the sensitivity that strong lines would show alone. It is comforting
to notice that the standard deviation of the $\xi$ and log($g$) values are in the
order of 0.2 km s$^{-1}$ and 0.2 dex respectively, as expected from the
procedure used in Sec. \ref{stellarparameters}. More interestingly, the
resulting dependence of the abundances on the EW uncertainties and, in turn, on
these two imperfectly determined parameters, is very small, of the order 0.02
dex for [Fe/H] and 0.04--0.06 dex for [Mg/H]. The real impact of the parameter
determination is thus much smaller than Table \ref{tabledependence} suggested.
This procedure is also useful for evaluating how these errors cancel out in the
ratio of the two elements, since we are interested in determining the [Mg/Fe]
ratio; the uncertainty varies between 0.03 dex and 0.06 dex for stars of higher and
lower metallicity. We re-emphasize that we did not derive
$T_{\rm eff}$ spectroscopically in this process, because of the lack of lines of
high EP. Consequently, an uncertainty from the adopted $T_{\rm eff}$ must still
be included, which we take from Table \ref{tabledependence}. However, this
uncertainty almost cancels out for the [Mg/Fe] ratio.

\subsection{Non-LTE effects on the abundances}

We derive chemical abundances assuming local thermodynamical equilibrium (LTE).
This assumption may be unrealistic, and one should be careful that lines may form
in conditions that depart from LTE. However, the current efforts to remove the
LTE assumption yield conflicting results, as shown by the results of
Th$\acute{\rm e}$venin \& Idiart (\cite{TI99}) and Gratton et {\it al.}
(\cite{Gratton99}) for Fe, and Zhao {\it et al.} (\cite{Zhao98}) for Mg I.

Non-LTE (NLTE) calculations from Th$\acute{\rm e}$venin \& Idiart (\cite{TI99})
show that iron abundances derived from Fe I lines should be increased by 0.3
dex, compared to the LTE estimates, for [Fe/H] in our metallicity range. On the other
hand, Mg abundances should be reduced by 0.04 dex when derived from the Mg
I line at 5528 \AA\ and by 0.03 dex for the two Mg Ib lines (Zhao {\it et al.}
\cite{Zhao98}). This would lead to a systematic overestimate of [Mg/Fe] of 0.33
dex in LTE (with a 0.04 dex difference in the NLTE corrections between the
lowest and the highest metallicity of our stellar sample). However, since Fe is
mostly present in its ionized state, Fe abundances derived from Fe II lines are
insensitive to overionization, and a higher value of log($g$)$_{NLTE}$ (by
0.3--0.4 dex following Th$\acute{\rm e}$venin \& Idiart (\cite{TI99})) would be
required to obtain ionization balance. This would further increase the
overestimate of [Mg/Fe] to 0.45 dex in LTE. 

It is important to note that Gratton {\it et al.}
(\cite{Gratton99}) found different results when calculating NLTE corrections to
Fe and Mg abundances. They suggest corrections that are very small, if not
negligible, for main-sequence and RGB stars. This is in conflict with the
calculations of Th$\acute{\rm e}$venin \& Idiart (\cite{TI99}), leaving NLTE
corrections still very uncertain.

The assumption of LTE is thus presumably introducing errors in our abundances
but, because of the homogeneity of our sample and the small variation of NLTE
corrections on the metallicity range considered, we assume any effect on
[Mg/Fe] is buried in our random uncertainties. Furthermore, because of the
present uncertainties on the effects of NLTE, and in order to have results
comparable with most observational studies, we adopt the LTE estimates in our
discussion, warning the reader that our [Mg/Fe] ratio may be overestimated.

\subsection{Abundance error estimates}

In Sec. \ref{stellarparameters} we estimated the uncertainties on the adopted
stellar parameters. In Table \ref{tabledependence} we showed the dependence of
the abundances on these parameters taken one at time; in Table
\ref{tablerandom} the results of the Monte Carlo test showed the combined
dependence on iteratively determined log($g$) and $\xi$ values to have a smaller
impact on the abundances than the quadratic sum of the two errors taken
separately. Moreover, the Monte Carlo test was performed propagating numerically
the uncertainty on the EW, thus including the contribution of the latter in the
result. We can now estimate the errors on our final abundances. We treat
separately the random errors that affect the scatter of our results, and the systematic
errors that affect the absolute values (and thus the average [Mg/Fe]).

\subsubsection{Random errors}

As shown in Table \ref{tablerandom}, the uncertainty on the EW propagates
through the determination of the parameters log($g$) and $\xi$, and affects the
resulting [Fe/H] by 0.02 dex, [Mg/H] by up to 0.06 dex, and [Mg/Fe] by up to
0.06 dex. We could adopt these values as the uncertainties on the abundances due
to the combination of the three factors, 1$\sigma_A(\sigma_{EW},\sigma_{{\rm log}(g)},
\sigma_{\xi})$. However, we make use of log($g$) values from the photometric
calibration, hence the Monte Carlo test does not reflect the entire procedure.
We are obliged to estimate our errors as the quadratic sum of the uncertainty
due to single parameter variations, and consider the results of the Monte Carlo
test as a conservative estimate of the error due to uncertainty on line
measurement, $\sigma_A(\sigma_{EW})$.

Adding the random uncertainty on temperature, $\xi$ and log($g$) (Table
\ref{tabledependence}) increases the uncertainty on the abundances to 0.04 dex
for [Fe/H], 0.10 dex for [Mg/H]$_{\rm Ib}$ and 0.08 dex for [Mg/Fe]$_{\rm Ib}$.
Stars where the weak Mg 5528 \AA\ line was not measured have a slightly
higher uncertainty on Mg, by about 0.01 dex.

Log (gf) values have been chosen on the basis of an experimental determination
with uncertainties $<$25\%. This implies that we have to account for a
corresponding $<$ 0.1 dex uncertainty in the abundances derived from single
lines. Having a number of lines $N$ will reduce the uncertainty in the average
by a factor 1/$\sqrt{N}$. For Fe abundances, where the number of lines is
statistically significant, we assume the standard error of the Fe abundance
averaged on all the lines to be a good estimate of the total random error. This
is up to 0.03 dex. For Mg, the two or three lines give a random error due to
$gf$-values of $\sim$ 0.06 dex. However, in the case of Mg, we treat these
errors as systematic, as described in the next section. Adding these
uncertainties quadratically to the previous estimates produces a total random
error of 0.05 dex for [Fe/H], 0.10 dex for [Mg/H] and 0.08 dex for [Mg/Fe]. 

A final remark is relevant on the uncertainties derived by RNB on their
temperature estimates. Since their 40K random uncertainty is very low, and our
[Fe/H] abundances are very sensitive to temperature, we explored the effect of
adopting an artificially more conservative $\sigma$ = 80K error for this study.
This would increase the random uncertainty on [Fe/H] to 0.07 dex and that on [Mg/H]
to 0.11 dex. The [Mg/Fe] error remains unchanged at 0.08 dex.

\subsubsection{Systematic errors}

We estimate the effect of systematic errors in our model parameters as follows.
We consider the effect of an overestimate of 150 K as the systematic error on
the temperature, based on the previous discussion of the photometric estimate
adopted in this study --- see Sec. \ref{stellarparameters}. The variation on
[Mg/Fe] for a systematic change of $-150$ K alone would be less than $-0.01$
dex, because of an equivalent decrease of $\sim$ 0.14 dex in both [Fe/H] and
[Mg/H]. However, the procedure used to determine the spectroscopic log($g$) would
partially compensate for the T$_{\rm eff}$ change, resulting in a decrease in
log($g$) of about 0.3 dex, leading to an overall increase of 0.04 dex for [Mg/H].
The overall variation on [Fe/H] is 0.12 dex, because of the much weaker
sensitivity to changes in log($g$). This produces a 0.08 dex change in [Mg/Fe],
because of the greater sensitivity of Mg Ib lines to log($g$). 

The uncertainty on $gf$-values can be a large contribution to our uncertainty on
the abundances, and requires further comment. For Fe, we introduced the
uncertainties on the $gf$-values in the random error. However, in the case of
the Mg Ib lines, the uncertainty on the $gf$-values should be treated like a
systematic effect, since the two Mg Ib lines are averaged only with the weak Mg
line at 5528 m\AA. Because of the similar EPs and strength of
the two Mg Ib lines, errors in the Mg $gf$-values could cause the same variation
in the line abundances of both, thus shifting our Mg abundances by the same
amount. This will affect the mean value of the Mg abundance of a star but not
the line-to-line scatter about the mean. We prefer to include the uncertainty in
the Mg $gf$-values separately from the random error derived in the previous
subsection. If we do so, the random error on [Mg/Fe] is 0.08 dex, while the
systematic error becomes the sum of the systematic error due to temperature
(0.08 dex) and the one due to the $gf$-values (0.06 dex), i.e. the total systematic
uncertainty on [Mg/Fe] is then 0.10 dex. The total systematic error for [Mg/H]
is 0.07 dex, while [Fe/H] is affected systematically only by the temperature, i.e. by
0.12 dex.    

\subsection{Metallicity comparison}

\begin{figure}
   \centering
   \includegraphics[width=88mm]{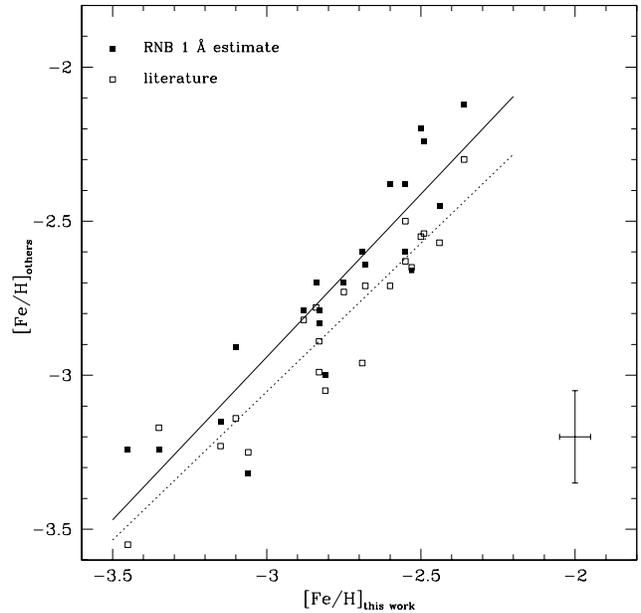}
      \caption{ A comparison between the Fe estimates of this study and the 1 \AA\ (filled squares) and literature (empty squares) estimates respectively. Overplotted are the two linear fits.  
              }
         \label{fecomparison}
   \end{figure}

Figure \ref{fecomparison} shows a comparison of the metallicity estimates of
this study with the 1 \AA\ estimates of RNB (filled dots). The 1\AA\ estimates
were obtained by a calibration of the pseudo-equivalent width of the Ca II K line
as a function of B--V colour to indicate the stellar metallicity [Fe/H] (Beers et
al. \cite{Beers99}). A linear fit $ax+b$ gives $a= 1.05$ and $b=0.23$, with an
rms of 0.14 dex, larger than our uncertainties on Fe (0.05), but consistent with
the value of 0.15 dex claimed by RNB for the 1 \AA\ estimates. In addition, we show
the trend of Fe literature values as discussed by RNB (empty squares). A linear
fit gives $a=0.96$ and $b=-0.16$, with a smaller scatter about the fit, rms=0.11
dex. This implies a slightly better agreement of our metallicity determination
with previous literature determinations than with the 1 \AA\ estimates. Both,
however, can be considered in good agreement with our estimates within the stated
uncertainties.

%

\section{Discussion}\label{discussion}

\subsection{Fe and Mg abundances } 

\begin{figure*}
   \centering
   \includegraphics[width=145mm]{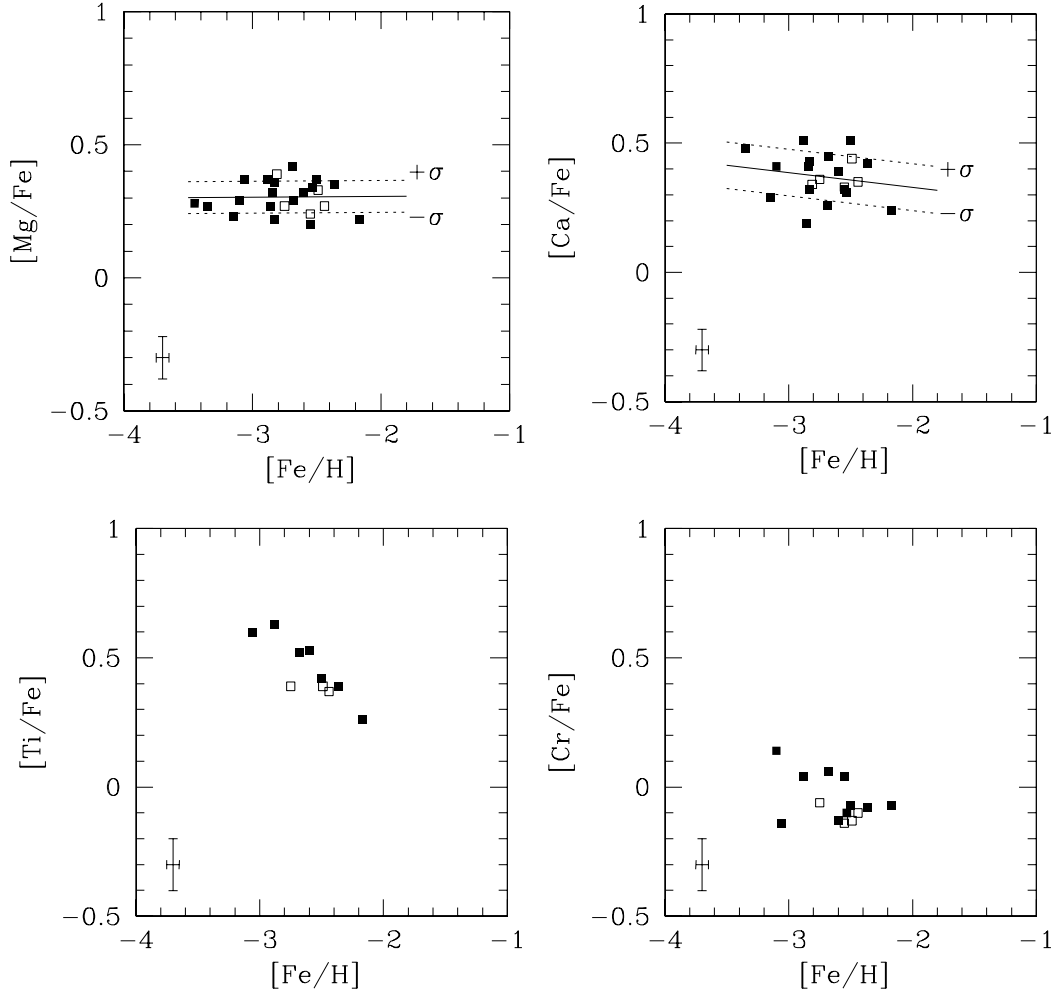}
   \caption{Abundances are shown as [el/Fe] vs [Fe/H]. Empty squares are stars with no Fe II measurements. In the case of Mg and Ca, we show the results of a linear fit (thick line) and the $\pm 1\sigma$ region. This is not shown for Ti and Cr because of the small number of stars with Ti and Cr measurements.}
              \label{mgfe}\label{plots}
    \end{figure*}

The abundances of Fe and Mg are shown in Table \ref{tableabundances}. 
[Mg/Fe] is within a range of 0.20 dex, between 0.22 and 0.42. Figure
\ref{mgfe} displays the run of [Mg/Fe] vs. [Fe/H]. The thick line is the result
of a linear least squares fit and the two thin lines delimit a $\pm 1\sigma$
region, $\sigma$ = 0.06 dex being the standard deviation of points about the
fit. The equation of the linear fit is:

\begin{equation}
{\rm [Mg/Fe]} = (0.00 \ \pm 0.04) \cdot {\rm [Fe/H]} + (0.31 \ \pm 0.12)
\end{equation}

The result of this linear fit suggests that the trend of [Mg/Fe] as a
function of [Fe/H] is flat.
We calculate a weighted mean
for [Mg/Fe], the weights being the inverse of the square of the adopted
uncertainties. The weighted average, standard error and standard deviation are

\begin{equation}
\overline{\rm [Mg/Fe]} = 0.30 \pm 0.01 \ \ \ \ \sigma = 0.06 
\end{equation}

Systematic errors are not accounted for in this statement;
they may affect the mean value by up to 0.10 dex, which we take to be the limit on the
precision with which the average value of [Mg/Fe] is determined. 

The scatter is not affected by systematic errors, hence, if no real intrinsic scatter was present,
we would expect the rms scatter (0.06 dex) to be no larger than the
random uncertainty (0.08 dex), as is the case. Moreover, the standard deviation of our results is similar to that of the Monte Carlo simulation
that was performed starting from the data of a single star (but which excluded
random errors in $T_{\rm eff}$). We thus infer that the
dispersion in the results is due to our analysis; any intrinsic
scatter in [Mg/Fe] of the sample must be much smaller, $\ll 0.06$ dex. 
This is our most important result, and is a considerable reduction on
the Cayrel {\it et al.} standard deviation on [Mg/Fe], 0.13 dex.
We consider the implications in Sec. \ref{implications}. 

The very low scatter confirms that our selection criteria and analysis produced a very homogeneous sample. A higher scatter could have
been the signature of an intrinsic characteristic of the stellar sample or star-to-star differences introduced by the analysis. The
latter can be difficult to estimate because of its dependence on the
adopted model atmospheres. As an example of the difficulties in quantifying the
real scatter, Johnson's (\cite{Johnson02}) study of giants
has a dependence of [Mg/Fe] on both log($g$) and $T$, [Mg/Fe] showing a very low
scatter in the run versus each parameter but a high scatter versus [Fe/H]. A
similar problem is buried in the results of Cayrel {\it et al.}
(\cite{Cayrel03}) whose mean, [Mg/Fe] = +0.27 dex, is in reasonable agreement with our
findings, but their standard deviation, 0.13 dex, has a component due to a slight
dependence of their abundances on temperature and gravity. Our utilisation of
turnoff stars covering only a narrow range of temperatures and surface gravity has avoided
this difficulty.  In less homogeneous studies, different stellar evolutionary
stages are considered in the same sample, thus including a star-to-star
difference in the dependence on the adopted stellar atmospheres. Finally, the
use of similar absorption lines for all the stars has limited the influences of
uncertainties in $gf$-values to a systematic uncertainty on the average [Mg/Fe],
not affecting the scatter. Further studies of larger, critically-selected samples will aid
in determining the intrinsic properties of the Galaxy. 

\subsection{Other elements} 

The abundances derived for the other elements are shown in Table
\ref{tableothers} based on up to five lines of CaI, three of Ti I
and three of Cr I. The trends of these abundances are shown in
Fig \ref{plots}.

For [Ca/Fe], we show a linear least square fits as a function of [Fe/H].  A statistical test indicates that the slope is not significant. Since we have Ca estimates for all
but two stars of our sample (CD--24$^{\circ}$17504 and LP831--70), we use
[Ca/Fe] to verify the consistency of our $\alpha$-element abundances. The mean
[Ca/Fe] is +0.37, with a slightly higher standard deviation than for [Mg/Fe],
0.09 dex. The estimated random error on [Ca/Fe] is slightly lower than the one on [Mg/Fe]
-- on average 0.07 dex. (The Monte Carlo simulation, which excludes the
temperature and the photometric calibration for the gravity, showed a much
smaller error for [Ca/Fe], of order 0.02 dex, because of the smaller sensitivity
of the weak lines used for Ca to changes in stellar parameters.) Consequently,
we cannot rule out the existence of an intrinsic scatter in the [Ca/Fe] values
up to $\sigma \sim$ 0.06 dex. 

Previous authors have used an $\alpha$-index averaging Mg and Ca chemical
abundances (see e.g., Gratton {\it et
al.} \cite{Gratton2000}). However, the Thielemann {\it et al.}
(\cite{Thielemann96}) SN II models show that Ca yields are
only slightly dependent on progenitor mass, whereas Mg yields depend quite strognly on progenitor mass. This leads iGCE
models to predict a lower scatter for Ca than for Mg (Argast {\it et al.}
\cite{Argast00}). Moreover, Ca is synthesised deeper in the progenitor than Mg,
so Ca yields are dependent also on the energy of the explosion, and hence not
as independent as possible of SN II model
uncertainties. We therefore concentrate on [Mg/Fe].

We also tabulate abundances for Cr and Ti. Cr absolute abundances are, on average, similar to those of Fe. We confirm the well-known overabundance of  Ti relative to Fe, though with a suspicious dependence on [Fe/H]. However, the
Ti abundances have been poorly determined due to the use of only one line for
most of the stars, and hence we place little weight on this result. We do not
investigate this element further.

\subsection{Implications for Galactic chemical evolution}\label{implications} 

\subsubsection{Restatement of the problem}

The absence of
intrinsic scatter in the Mg abundances presents a real challenge for
inhomogeneous Galactic chemical evolution models. 
Argast {\it et al.} (\cite{Argast00}, \cite{Argast02}) and Tsujimoto \& Shigeyama
(\cite{Tsujimoto98})
have shown that iGCE models predict a large scatter in [Mg/Fe] at [Fe/H]$<-2.5$ that reflects the dependence of 
nucleosynthesis on the progenitor mass of individual SN II, at a time in the history of GCE when the ISM was poorly mixed.
The Fran\c{c}ois {\it et al.} (\cite{Francois04}) homogeneous GCE model can
reproduce the abundance patterns observed in the most recent observations by
requiring a large correction to SN II yields. One has to investigate why
homogeneity occurs in a metallicity region expected to be inhomogeneous. 

The Argast {\it et al.} model predicts that for [Fe/H]$ < -3.0$, SN II pollute
only locally, i.e., the ISM is mixed at a rate slower than successive SN II events, and individual SN yields are reflected by the local ISM. In the range $-3.0<$ [Fe/H] $<-2.0$, mixing of the ISM causes the first overlaps of polluted
regions, though some regions that are unenriched remain. At [Fe/H]$=-2.5$,  some regions of the ISM are enriched several times
by supernova progenitors of different masses. One should thus observe stars
showing abundance patterns characteristic of SN progenitors
of different masses. By [Fe/H] $> -2.0$, the yields from successive polluting
events are averaged over the initial mass function (IMF) and well mixed, bringing iGCE
and 1-zone GCE models into agreement. Our metallicity range
($-3.4<$[Fe/H]$-2.2$) covers the inhomogeneous part of the ISM history, where a
range up to 1 dex is predicted for Mg abundances, compared to a range of 0.2 dex for our observations. 

\begin{figure}
   \centering
   \includegraphics[width=88mm]{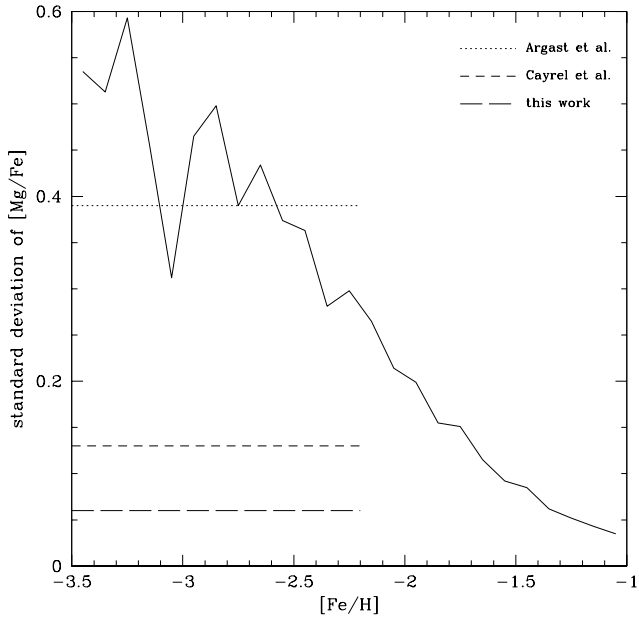}
      \caption{ The standard deviation of the dispersion in [Mg/Fe] predicted by the model of Argast {\it et al.}  (\cite{Argast02} their Fig. 1), in the case of SN II yields adopted from Thielemann {\it et al.}  (\cite{Thielemann96}). The dotted, small dashed and the dashed lines show the standard deviation over the metallicity range of our observations, $-3.4<$[Fe/H]$<-2.2$, respectively for Argast {\it et al.}  (\cite{Argast02}), Cayrel {\it et al.} (\cite{Cayrel03}) and this work.       
              }
         \label{argast1}
   \end{figure}

\begin{figure}
   \centering
   \includegraphics[width=88mm]{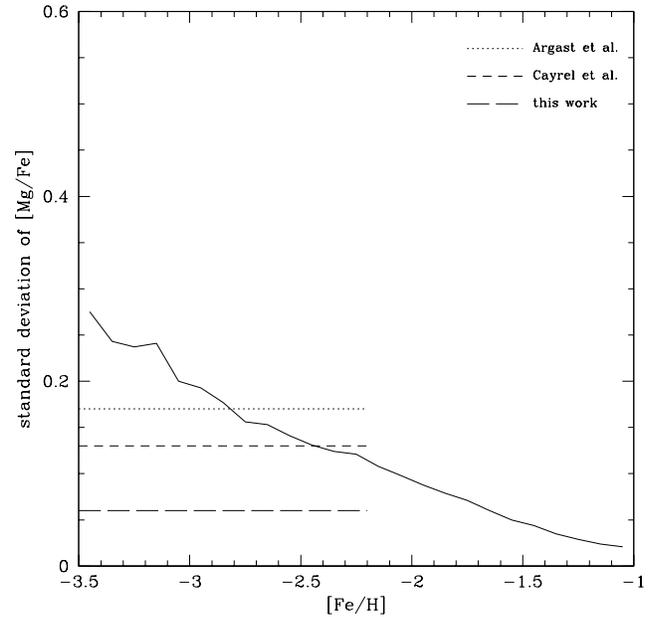}
      \caption{ The standard deviation of the dispersion in [Mg/Fe] predicted by the model of Argast {\it et al.}  (\cite{Argast02} their Fig. 14), in the case of their H1 model. The dotted, small dashed and the dashed lines show the standard deviation over the metallicity range of our observations, $-3.4<$[Fe/H]$<-2.2$, respectively for Argast {\it et al.}  (\cite{Argast02}), Cayrel {\it et al.} (\cite{Cayrel03}) and this work.   
              }
         \label{argast2}
   \end{figure}

In Figs. \ref{argast1} and \ref{argast2}, we show the standard deviation of
Argast's model calculated for 0.1 dex metallicity bins. The two figures
correspond to the models shown in Figs. 1 and 14 of Argast {\it et al.} (\cite{Argast02}),
calculated adopting SN II yields from Thielemann {\it et al.}
(\cite{Thielemann96}), 
and from the Argast {\it et al.} 'H1' empirical model,
respectively. 
The yields from Thielemann {\it et al.} (\cite{Thielemann96})
and Nomoto {\it et al.} (\cite{Nomoto97}), adopted in the
Argast {\it et al.} model, range from [Mg/Fe]=$-0.83$ for a
progenitor of 13 M$_{\odot}$ to [Mg/Fe]=1.44 for a progenitor of 70 M$_{\odot}$,
with a monotonic relationship between the Mg mass yield and the progenitor mass.
The H1 model was obtained by modifying the SN II
yields to match the Fe yields inferred from SN II observations, rather than
using yields from a SN II model, and was further constrained by
the range of abundances previously observed in low-metallicity stars. This model shows a much lower scatter due to a smaller adopted range in SN~II Mg/Fe yield ratios. 
Superimposed on these figures are shown our standard
deviation for [Mg/Fe], and the predicted mean standard deviations of the models  calculated over the same
metallicity range. We also show the standard deviation of [Mg/Fe] from Cayrel
{\it et al.} (\cite{Cayrel03}). 
For the H1 model, the predicted standard deviation in [Mg/Fe] is $\sim$0.17 dex, rising towards lower metallicities, and
well in excess of our limit of 0.06 dex. 

Theoretically (Arnett \cite{Arnett}), the assumption of instantaneous mixing
adopted by 1-zone GCE models is not correct at these low metallicities, hence
inhomogeneity must be considered in the models. 
If iGCE models are correct in relaxing the instantaneous mixing assumption, then
the lack of scatter in the [Mg/Fe] values we have measured is all the more
remarkable. To explain this result, one has to focus on SN II yields and other
uncertainties in the models. 

%

%

\subsubsection{A restricted mass range for SN II progenitors}

Chiappini {\it et al.} (\cite{Chiappini99})
found that, for metallicities $-3.0<$[Fe/H]$<-1.0$, the trends in $\alpha$-elements are expected to show slopes. This is not seen in our flat trend at [Fe/H] $<
-2.2$. However, this slope is not predicted in other models (e.g. Argast
\cite{Argast02}). The uncertainty on our measurement of the slope restricts the
variation of the [Mg/Fe] slope to within 0.04 dex from flat over
the metallicity range $-3.0<$[Fe/H]$<-2.0$.

The range of progenitor masses that give rise to
[Mg/Fe]=0.3$\pm 0.1$ (our observational value) is 18--20 M$_{\odot}$
according to the yields of Thielemann {\it et al.} (\cite{Thielemann96}).
However, even if we knew the exact yields of this progenitor mass range, the
flat trend and the lack of intrinsic scatter raises the question:  Why, at any time in the early
Galaxy, would the only contribution to the ISM chemical composition of Mg come
from this narrow range of progenitors? 

The upper mass limit that avoids complete collapse to a black hole is unknown,
and while it is believed to be in the range of 30--50 M$_{\odot}$, there is some
possibility it could be lower in low-metallicity stars. Maeder (\cite{Maeder92})
found that a limit around
20--25M$_{\odot}$ was consistent with $\Delta Y/\Delta Z$ (the
enrichment of helium relative to metals). More recently, Fryer (\cite{Fryer99}) confirmed the 20 M$_{\odot}$ limit for the initial mass for black hole formation. Fryer and Heger (\cite{FHeger00}) suggested that stellar wind and rotation could decrease the energy of the explosion and thus the lower mass limit for black hole formation. 

However, even if the upper mass limit for Mg enrichment of the ISM could
be 20 M$_{\odot}$, there is no reason to think that progenitors
as low as 10--12 M$_{\odot}$ would be disabled. One may question our theoretical understanding
of SN II models, or look for criteria that would favour SN II with this
progenitor mass. One possibility is a shallower (top-heavy) IMF in the early phases to
raise the production of massive stars, allowing progenitors near the black-hole mass limit (20 M$_{\odot}$?) to dominate enrichment of
the ISM. 10 M$_{\odot}$ SN II progenitors would nevertheless still evolve within $\sim$ 30 Myr, which is the
timescale for the enrichment of the halo to [Fe/H]=--3.0 according to models
which explain the rise of the r-process enrichment in [Ba/Fe] and [Eu/Fe] in the halo in terms of low mass
(8--10 M$_{\odot}$) SN II (Mathews \& Cowan \cite{MC90}; Travaglio {\it et
al.} \cite{Travaglio01}; Ishimaru {\it et al.} \cite{I04}). 

On balance, we conclude that currently there is no support for the proposition that only a narrow range of SN II progenitor masses around 18--20
M$_{\odot}$ contributes significantly to the Galactic enrichment in Mg. 

\subsubsection{SN II yields}

The iron yields from SN II are notoriously difficult to predict (Nomoto {\it et
al.} \cite{Nomoto97}); Thielemann {\it et al.}
(\cite{Thielemann96}) assume a decline of Fe with rising progenitor mass, while
Woosley \& Weaver (\cite{WW95}) assume an increase. 
Given this uncertainty, could the majority of SN II produce the same [Mg/Fe] ratio in
their ejecta? A Mg/Fe yield ratio constant with progenitor mass (and thus constant over the history of the ISM) would require a proportionality between the size of the ejected Si-burned shell (the
source of Fe) and the mass of the progenitor (which correlates with Mg yield). It seems highly
unlikely that nature should achieve such a perfect balance in the hydrostatic
production of Mg and the subsequent ejection of a fraction of the explosive production of Fe,
across the $\sim$ 10-20 or 10-30 M$_{\odot}$ mass range of SN II progenitors. Moreover,
the flat slope found for [Ca/Fe] versus metallicity would be affected by a
change in the Fe yields, as would other element [el/Fe] ratios, and the Ca yield would also depend on explosion energy and progenitor
mass. Thielemann's models do not show a
variation of Ca with progenitor mass of the same order as that of Mg. 

Attempts
to reproduce the chemical pattern of very low-metallicity
stars by adjusting the mass cut in SN II explosion models have been unsuccessful (Chieffi \& Limongi \cite{CL02}). Tsujimoto \& Shigeyama
(\cite{Tsujimoto98}) traced yield versus mass for SN II from the
observations of McWilliam {\it et al.} (\cite{Mcwilliam95}) and the yields from
Woosley \& Weaver (\cite{WW95}), and satisfactorily produced predictions for
averaged abundances ratios within 0.1 dex of the observations. Their final
remark is the need of revising SN II yields of elements except C, O, and Mg. Our
results strongly indicate that either also Mg yields are wrong (see also Argast
{\it et al.} \cite{Argast02}), and Fe yields are wrong, or the theoretical
framework of iGCE has reached its limit and needs a deeper understanding of 
mixing. 

Fran\c{c}ois {\it et al.} (\cite{Francois04}) have shown that constraints to SN II yields can be found by matching observations with their homogeneous GCE model.
However, even if SN II yields are revised following their prescription, within the framework
of homogeneous models, the lack of scatter still remains to be explained. If [Mg/Fe]
yields truly are not constant for all progenitor masses, and there is not only a very narrow mass range of SN II progenitors that produce significant Mg, then the lack of
scatter points to the following unexpected conclusion: the ISM
from which stars formed at [Fe/H] $>-3.5$ was on the whole already well-mixed, {\it and} it was enriched by a 
sufficient number of SN II to show an IMF averaged pattern at [Fe/H]$<-3.0$, in spite
of the expectations to the contrary from iGCE current models. Furthermore, the hypothesis by
Cayrel {\it et al.} (\cite{Cayrel03}) of the existence of a plateau in the
[el/Fe] ratios at very low metallicity is clearly supported by our [Mg/Fe]
ratio.  We are left to puzzle over
how this IMF average is achieved ove much of the halo on the short timescales ($\leq$ 30 Myr) for the
ejection of the full range of SN II progenitor masses (M $\geq$ 10M$_{\odot}$). This could imply that we are observing the primordial abundances from the
first stars below [Fe/H]=$-3.0$, as Cayrel et al. suggested.

\subsubsection{Mixing and cooling timescales}

Enrichment timescales are not well constrained.
Prantzos (\cite{Prantzos03}) shows that an early
phase of infall and relaxing the instantaneous recycling
approximation causes his halo outflow model to reproduce better the metallicity
distribution of stars in the halo. Enrichment occurs much quicker in his model than in iGCE
models, but the introduction of an early infall phase in his model slows the
metallicity enrichment again, as a result of which the time needed to reach [Fe/H]=$-3.0$ increases from 35
Myr to 100 Myr. A similar delay in the enrichment could occur in iGCE models for a strong early infall phase. However, a slower enrichment of the Galaxy does not affect
the predicted scatter in [Mg/Fe] vs. [Fe/H], but rather the age-metallicity
relation. 

De Avillez \& Mac Low
(\cite{AM02}) studied the mixing timescale in a SNe-driven ISM, and
show that inhomogeneities due to single SN II explosions take up to 350 Myr to
be erased (for the present Galactic SN II rate). Even when mixing
scales as small as a few kpc are considered, efficient mixing requires of
order 120 Myr. Increasing the SN rate by a factor 10, they find that inhomogeneities take only some tens of Myr to disappear. If increasing
the SN II rate reduces the time needed for efficient mixing, then for a short
time ($\sim$ 30Myr) it favours the explosion of higher mass progenitors (because
of their shorter lifetimes) and a higher average [Mg/Fe] ratio.
This effect could also be achieved with a top-heavy IMF (discussed above).
This would avoid the ISM having regions with low [Mg/Fe] values, but would still require a progenitor upper mass limit of 20 M$_{\odot}$ to avoid the production of [Mg/Fe] $>$ 0.4 except for at the very lowest metallicities (Norris (\cite{Norris04})). 

We asked above how an IMF-averaged [Mg/Fe] value could be achieved on the very
short evolutionary timescales of stars at [Fe/H]$<-3.0$. This may be the wrong
question to ask if {\it cooling} timescales crucial to star formation, rather than {\it mixing} timescales, are the
critical factor. iGCE models lack reliable treatments of the cooling 
of hot SN II ejecta in a metal-poor environment. If the cooling time exceeds the
evolutionary time of the lowest mass SN~II progenitors of 10 M$_{\odot}$, i.e. 30 Myr, then a complete
IMF-average of the [Mg/Fe] ratio will be achieved, not on the evolutionary time
of the SN II, but on the cooling timescale of the ISM, and therefore on the
formation timescale of the next stellar generation.
 
Oey (\cite{Oey03}) investigates mixing and cooling processes in the ISM by means
of her simple inhomogeneous model (SIM) of Galactic chemical evolution, which
explicitly incorporates interstellar mixing and mass transport. As previously
found by De Avillez \& Mac Low (\cite{AM02}), diffusion is inefficient compared
to turbulent mixing. Even though ``turbulent processes are extremely difficult to
constrain'', turbulent mixing is found to be extremely efficient in the case of a
hot ionized medium (HIM). Thus, in a phase of HIM-dominated ISM, efficient
mixing could indeed occur. Furthermore, a hot ISM has a low cooling efficiency,
thus a two-phase ISM could experience a delay due to
the cooling time between the SN II events and the incorporation of their ejecta in
star forming regions. Once new stars could form from SN-enriched gas, they would form out of a well-
mixed ISM. By a comparison with the observed metallicity distribution function and the lowest metallicity observed in stars ([Fe/H]=--4.0 at the time of Oey's analysis),
Oey's points out that an inhomogeneous phase could have been
extremely short lived and that the ISM would have then mixed efficiently. Furthermore, Recchi {\it et al.} (\cite{Recchi01}) find that, in the case of dwarf galaxies, a single sturbust leads to the development of galactic winds and to a quick mixing. However, dwarf galaxies evolve in a low gravitational potential and thus dynamical processes (such as galactic wind) may have larger influences on the mixing than in the case of the Galaxy.
 

We believe further investigations of mixing and cooling timescales in relation
to iGCE models are advisable, since there is no clear theoretical expectation
that the problem of the missing inhomogeneities will be solved by adjustments of
the SN II yields, the SN II progenitor mass range, or modification of the IMF.     

\section{Summary}\label{summary}

We have presented the results of the chemical analysis of a sample of 23
main-sequence turnoff halo stars previously selected by RNB. Because of the small number
of lines identified in the most metal-poor stars, the [Mg/Fe] abundance ratio
was determined on the basis of photometric calibrations for both effective
temperature and surface gravity. This method should avoid introducing
star-to-star differences and preserve the homogeneity of sample. 

Our results show that [Mg/Fe] is constant over the metallicity range considered.
There is no indication of internal scatter, the 0.06 dex standard deviation
about the mean being well within the expected 0.08 dex random uncertainty. The
maximum point-to-point range of [Mg/Fe] is only 0.2 dex.

The observed lack of scatter presents a challenge for iGCE models that predict
a dispersion for [Mg/Fe] at low metallicity, with standard deviation of the
order 0.4 dex, and a total range as large as 1.0 dex. We review the possible
sources of uncertainties in the iGCE models, such as the mass range of SN II
progenitors and the Mg, Ca and Fe yields, and the IMF, but find no obvious
solution to the problem. We identify the need for exploring the time dependence
of cooling and mixing in the early Galaxy, to verify whether these processes
could cancel out the predicted scatter by ensuring that new stars form only once
the enrichment of the ISM has been averaged over the IMF for the full range of
SN II progenitor masses. This could occur during a hot early phase of the ISM,
where turbulent mixing is extremely efficient in slowly cooling, hot,
metal-deficient gas.  

It is still difficult to reconcile the large spread in [Sr/Fe] at
[Fe/H] $<$ --2.0 (discussed in Section 1) with the lack of a spread in
[Mg/Fe] found in Section 4.  The latter argues for star formation
occurring only after an IMF average of Mg yields has been achieved in
the ISM, whereas the lack of uniformity in [Sr/Fe] requires the
existence of stochastic events that are not completely averaged out.
If the source of Sr is more restricted in mass, or otherwise rarer
than the source of Mg (which comes from the entire SN II range), then
this apparent inconsistency might be resolved.  The rise in [Ba/Fe]
toward [Fe/H] = --2.5, which may be due to the origin of r-process
elements in 8--10~M$_{\odot}$ stars (Section 5.3.2) again argues
against an IMF average being achieved by [Fe/H] = --3.0.  However, if
the Mg yield of such low mass stars is almost insignificant, as theory
suggests, then the IMF average we infer from the Mg observations is
probably a constraint only on higher mass SN II progenitors.  (That is
to say, the IMF average claimed for Mg is an average over most of the
IMF at masses M $\sim$ 20~M$_{\odot}$ rather than an average over
strictly all of the SN II progenitor range.)  The different behaviour
of [Ba/Fe] and [Mg/Fe] emphasises, according to this interpretation,
that Ba and Mg are predominantly synthesized in different progenitor
mass ranges.
  

E.A. acknowledges Roberto Gallino for useful suggestions and discussions.

D.A. acknowledges funding by the Swiss National Science Foundation.

J.N.E. acknowledges funding by the Australia Research Council

T.C.B. acknowledges partial funding for this work from grants AST 00-98508 and
AST 00-98549, as well as from grant PHY 02-16783: Physics Frontiers Center/Joint
Institute for Nuclear Astrophysics (JINA), awarded by the U.S. National Science
Foundation.


\end{document}